\documentclass[aps,prd,preprint,groupedaddress,longbibliography,nofootinbib]{revtex4-1}

\usepackage[utf8]{inputenc}
\usepackage{amsmath,amssymb}
\usepackage{amsfonts}
\usepackage{slashed}
\usepackage{xcolor}

\usepackage[pdftex]{hyperref}
\hypersetup{ 
pdfstartview=FitV,
colorlinks=true, 
linkcolor=black, 
citecolor=blue, 
filecolor=black, 
urlcolor=magenta,
}
\usepackage{graphicx}

\begin{document}

\preprint{CPHT-RR116.122018}
\preprint{IFT-UAM/CSIC-18-130}
\preprint{NORDITA 2018-125}

\title{Universal relaxation in a holographic metallic density wave phase}


\author{Andrea Amoretti}
\affiliation{Dipartimento di Fisica, Universit\`a di Genova,
via Dodecaneso 33, I-16146, Genova, Italy and I.N.F.N. - Sezione di Genova
}
\affiliation{Physique  Th\'{e}orique  et  Math\'{e}matique  and  International  Solvay  Institutes  Universit\'{e}  Libre de Bruxelles, C.P. 231, 1050 Brussels, Belgium}
\email{andrea.amoretti@ge.infn.it}
\author{Daniel Are\'an}
\affiliation{Instituto de F\'\i sica Te\'orica UAM/CSIC,
	Calle Nicol\'as Cabrera 13-15, Cantoblanco, 28049 Madrid, Spain}
\email{daniel.arean@uam.es}
\author{Blaise Gout\'eraux}
\affiliation{CPHT, CNRS, Ecole polytechnique, IP Paris, F-91128 Palaiseau, France}
\affiliation{Nordita, KTH Royal Institute of Technology and Stockholm University, Roslagstullsbacken 23, SE-106 91 Stockholm, Sweden}
\email{blaise.gouteraux@polytechnique.edu}
\author{and Daniele Musso}
\affiliation{Departamento  de  F\'{i}sica  de  Part\'{i}culas,  Universidade  de  Santiago  de  Compostela  and  Instituto  Galego  de  F\'{i}sica  de  Altas  Enerx\'{i}as  (IGFAE).}
\email{daniele.musso@usc.es}

\date{\today}

\begin{abstract}

In this work, we uncover a universal relaxation mechanism of pinned density waves, combining Gauge/Gravity duality and effective field theory techniques. Upon breaking translations spontaneously, new gapless collective modes emerge, the Nambu-Goldstone bosons of broken translations. When translations are also weakly broken (eg by disorder or lattice effects), these phonons are pinned with a mass $m$ and damped at a rate $\Omega$, which we explicitly compute. {This contribution to $\Omega$ is distinct from that of topological defects.} We show that $\Omega\simeq G m^2\Xi$, where $G$ is the shear modulus and $\Xi$ is related to a diffusivity of the purely spontaneous state. This result follows from the smallness of the bulk and shear moduli, as would be the case in a phase with fluctuating translational order. 
At low temperatures, the collective modes relax quickly into the heat current, so that late time transport is dominated by the thermal diffusivity. In this regime, the resistivity in our model is linear in temperature and the ac conductivity displays a significant rearranging of the degrees of freedom, as spectral weight is shifted from an off-axis, pinning peak to a Drude-like peak. These results could shed light on transport properties in cuprate high $T_c$ superconductors, where quantum critical behavior and translational order occur over large parts of the phase diagram and transport shows qualitatively similar features.

\end{abstract}


\maketitle
{
The spontaneous formation of charge density waves and other types of translational order in weakly-coupled one-dimensional metals is well-understood: the Peierls instability opens a gap and a collective mode of electron-hole pairs is formed \cite{RevModPhys.60.1129}. After condensation, the spectrum of the system contains a gapless mode (the phase of the condensate, ie the Goldstone  boson generated by the spontaneous breaking of translations): physically, the density wave can slide without energy cost. In the presence of weak disorder or lattice effects, the Goldstone (or with a slight abuse of terminology, phonon) is pinned. More formally, weak explicit breaking of translations generates a small phonon mass $m$. 

Pinning has dramatic consequences on the linear frequency-dependent response. Assuming a normal metallic state with an ac conductivity characterized by a Drude-like peak centered at zero frequency, the spatially modulated state is characterized by a transfer of spectral weight to finite frequencies: an off-axis peak is formed at the so-called pinning frequency $\omega_o\sim m$. In classic treatments \cite{RevModPhys.60.1129}, the Goldstone contribution vanishes at zero frequency and the system is an electrical insulator.

The pinning peak is also characterized by its width, which heuristically captures the typical lifetime of excitations of the system. There are two well-known contributions to the peak width \cite{RevModPhys.60.1129}. One comes from the rate $\Gamma$ at which the momentum of the system relaxes due to disorder or lattice effects. The other comes from the rate $\Omega$ at which the Goldstone relaxes in the presence of mobile defects \cite{PhysRevLett.41.121}. In direct analogy to how vortices relax superfluid phase gradients and eventually destroy superfluidity, dislocations/disclinations also spoil phase coherence of the density wave. The phonon contributes to the dc conductivity $\sigma_{dc}\simeq \omega_p^2\Omega/(\Gamma \Omega+\omega_o^2)$ where $\omega_p$ is the plasma frequency (the Drude weight), \cite{Delacretaz:2017zxd}. As temperature decreases, defects freeze out, leading to vanishing phase relaxation and insulating behavior.

Density waves play an important role in the phase diagram of cuprate high $T_c$ superconductors. These materials are doped Mott insulators \cite{Keimer2015}, which tend to order translationally upon doping \cite{PhysRevB.40.7391,PhysRevB.54.7489,Kivelson1998,RevModPhys.75.1201,PhysRevB.83.104506,2018NatMa..17..697P,2018arXiv180904949A,2017Sci...358.1161H}. In the strange metallic phase, their resistivity scales linearly with temperature \cite{Bruin804} and violates the Mott-Ioffe-Regel limit \cite{RevModPhys.75.1085}. This signals the breakdown of the quasiparticle picture and is usually interpreted as a sign of strongly-coupled dynamics. This behavior has been attributed to an underlying metallic quantum critical point with maximal, Planckian dissipation  \cite{sachdev_2011,Zaanen2004} characteristic of strongly-coupled phases.

In the strange metallic phase, a shift of spectral weight is often observed with sharp Drude peaks at low temperatures moving off-axis as temperature increases \cite{2004PMag...84.2847H}. Quantum melting of stripe order by proliferating defects has been proposed as the origin of the strange metallic phase, see eg \cite{Kivelson1998,doi:10.1146/annurev-conmatphys-070909-104117}, but microscopic descriptions of fluctuating stripes in the vicinity of a quantum critical metallic phase and their effects on the ac conductivity remain a challenge \cite{PhysRevLett.84.5608,PhysRevB.82.075128,PhysRevLett.108.267001,PhysRevB.86.115138}. Based on an effective field theory approach (hydrodynamics), \cite{Delacretaz:2016ivq} argued that fluctuating translational order could cause the shift in spectral weight mentioned above. However the relaxation parameters $\Omega$, $\Gamma$ and $\omega_o$ enter as an input in the hydrodynamic theory, and need to be computed by a microscopic theory valid at strong coupling.

In this Letter, we combine Gauge/Gravity duality with hydrodynamics \cite{Delacretaz:2017zxd} to study the long distance transport properties of strongly-coupled, weakly-pinned density waves in the vicinity of a metallic quantum critical phase with a $T$-linear resistivity. Gauge/Gravity duality allows to address the strongly-coupled dynamics of certain quantum field theories by mapping them to a weakly-coupled theory of gravity \cite{Maldacena:1997re,Ammon:2015wua,Zaanen:2015oix,Hartnoll:2016apf}. 

We identify unambiguously a new contribution to the phase relaxation rate $\Omega$ due to weak explicit breaking of translations, which drives the shift in spectral weight and dominates the resistivity. While this type of damping is well-known from studies of the magnetic field-induced melting of Wigner solids \cite{PhysRevB.18.6245,PhysRevB.62.7553}, and phonon damping by disorder was studied very early on, see eg \cite{PhysRevB.17.535,PhysRevB.28.340}, it was reported to affect the ac conductivity only through the pinning frequency $\omega_o$ and the momentum relaxation rate $\Gamma$ \cite{RevModPhys.60.1129,Delacretaz:2017zxd}. Our results should also shed light on previous holographic studies of pinned density waves \cite{Ling:2014saa,Baggioli:2014roa,Jokela:2017ltu,Andrade:2017cnc,Alberte:2017cch,Andrade:2017ghg,Andrade:2018gqk}, where this contribution was not explicitly identified. In passing, we verify the validity of the hydrodynamic theory of damped and pinned density waves written in \cite{Delacretaz:2017zxd}.

Furthermore, we uncover a universal relation between the phonon damping rate and its mass, 
\begin{equation}
\label{Omegarel}
\Omega\simeq G m^2\Xi =\chi_{\pi\pi}\omega_o^2\Xi\quad \Rightarrow\quad\rho_{\textrm{dc}}\equiv\sigma_{dc}^{-1}\simeq\frac{1}{\rho^2\Xi}\,.
\end{equation}
$\Xi$ is a phonon diffusive transport coefficient, $G$ the phonon shear modulus, $\rho$ the charge density, $\chi_{\pi\pi}$ the momentum static susceptibility and the pinning frequency is defined from the phonon mass as $\omega_o^2\equiv Gm^2/\chi_{\pi\pi}$. It is universal since it determines microscopic parameters in terms of universal thermodynamic/hydrodynamic data. 
The expression for the resistivity  is reminiscent of an Einstein relation. We show that this relation is true in the limit of small bulk and shear moduli. We expect it will hold more generally in phases with fluctuating translational order, or close to a phase transition towards a translationally-ordered phase, where these quantities are small.}

Remarkably, at low temperatures, we observe that the system saturates a bound ensuring the positivity of entropy production. This is naturally explained by relaxation of the phonons into the heat current. The resistivity is then controlled by the thermal diffusivity $D_T\sim1/T$. Our result resonates with the idea that the thermalization time in strange metals and other strongly-coupled quantum phases is bounded from below by the `Planckian' timescale $\tau_P\sim\hbar/(k_B T)$ \cite{sachdev_2011,Zaanen2004} and that production of entropy is minimal eg through a lower bound on the shear viscosity \cite{Kovtun:2004de,Davison:2013txa,Zaanen:2018edk}.

We now explain in more detail how we arrive at these results, and close by commenting further on the relevance of our results to strange metals.

A number of technical details are relegated to a few Appendices.

\section{Holographic model}

We consider the holographic model \cite{Donos:2013eha} $S=\int d^{4}x\,\sqrt{-g}\mathcal L$
\begin{equation}\label{action}
\mathcal L=R-\frac12\partial\phi^2-V(\phi)-\frac14 Z(\phi)F^2-\frac12\sum_{I=1}^{2}Y(\phi)\partial \psi_I^2\,,
\end{equation}
with the scalar couplings behaving near the AdS boundary as $V_{uv}(\phi)=-6-\phi^2+O(\phi^3)$, $Z_{uv}(\phi)=1+O(\phi)$, $Y_{uv}(\phi)=\phi^2+O(\phi^3)$. For concreteness, in our numerical calculations we work with $V(\phi)=-6\cosh(\phi/\sqrt3)$, $Z(\phi)=\exp(-\phi/\sqrt3)$, $Y(\phi)=(1-\exp\phi)^2$. Details of our numerical procedure are given in appendix \ref{app:NumMeth}.
 
The model \eqref{action} enjoys a global shift symmetry $\psi_I\mapsto\psi_I+c_I$. Adopting the background Ansatz $\psi_I=k x^I$, $x^I=\{x,y\}$, the product of spacetime translations and shifts is broken to a diagonal $U(1)$. The $\psi_I$ then transform non-linearly under translations precisely as Goldstones of broken translations are expected to. The other background fields can consistently be taken to depend solely on the holographic radial coordinate, $ds^2=-D(r)dt^2+B(r) dr^2+C(r)(dx^2+dy^2)$, $\phi=\phi(r)$, $A=A(r)dt$.
Near the AdS boundary $r\to0$, $\phi(r)=\lambda r+\phi_{(v)} r^{2}+O(r^3)$, where $\lambda$ is the source and $\phi_{(v)}$ the vev of the operator dual to $\phi$. Moreover,
the bulk scalar fields can be rewritten as complex scalars $\Phi_I= \phi\exp(i\psi_I)$ close to the boundary. This shows that the breaking is spontaneous when $\lambda=0$ or explicit when $\lambda\neq0$ \cite{Amoretti:2016bxs,Amoretti:2017frz,Amoretti:2017axe}, which is the main focus of this work.

This holographic model does not capture the phase transition between the normal and the ordered phase. 
Instead, it describes the low energy dynamics of the (pseudo)phonons in the ordered phase, consistent with the pattern of symmetry breaking in an isotropic, two-dimensional Wigner crystal (WC) \footnote{A similar model was considered in \cite{Iqbal:2010eh} to study spin density waves. Previous recent holographic studies of collective transport in phases that break translations spontaneously include \cite{Jokela:2016xuy,Jokela:2017ltu,Alberte:2017cch,Andrade:2017cnc,Alberte:2017oqx,Andrade:2017ghg}.}.

 Finite temperature, finite density states are modeled by charged black holes in the bulk, which implies the existence of a regular Killing horizon at $r=r_h$. Hereafter, we use a subscript $h$ to denote quantities evaluated at $r=r_h$.
 
\section{Unrelaxed WC hydrodynamics from holography}

Aspects of unrelaxed WC hydrodynamics were studied previously in \cite{Amoretti:2017frz,Amoretti:2017axe}, and are reviewed in appendix \ref{app:WCreview}. The hydrodynamic retarded Green's functions at zero wavevector for the electric current and the phonons are \cite{chaikin_lubensky_1995,Delacretaz:2017zxd}
\begin{equation}
\label{KuboUnrelaxed}
G^R_{j j} = \frac{\rho^2}{\chi_{\pi\pi}}-i\omega\sigma_o\,,\quad G^R_{j\varphi}=-\gamma_1-\frac{\rho}{\chi_{\pi\pi}}\frac{i}\omega\,,\quad G^R_{\varphi\varphi}=\frac{1}{\chi_{\pi\pi}\omega^2}-\Xi\frac{i}{\omega}\,.
\end{equation}
$\sigma_o$, $\gamma_1$ and $\Xi$ are diffusive transport coefficients that appear in the longitudinal sector of hydrodynamic constitutive relations at first order in gradients \cite{Delacretaz:2017zxd}. $\chi_{\pi\pi}=sT+\mu\rho +k^2 I_Y$, with the entropy density $s$ given by the horizon area, $\rho$ by the electric flux out of the horizon and the chemical potential $\mu=A(0)$. The bulk and shear moduli are $2K=G=k^2 I_Y+O(k^4)=k^2\int_0^{r_h}\sqrt{BD}Y+O(k^4)$ \cite{Alberte:2017oqx,Amoretti:2017frz,Alberte:2017oqx}.
$\sigma_o$ has been computed in \cite{Amoretti:2017frz,Amoretti:2017axe,Donos:2018kkm}, and the computation of $\gamma_1$ and $\Xi$ is presented in {the companion paper \cite{Amoretti:2019}}:
\begin{equation}
\begin{split}
\label{gamma1xiUnrelaxed}
&\sigma_o=\frac{(sT+\mu\rho)^2}{ (\chi_{\pi\pi})^2} Z_h+\frac{4\pi\rho^2k^2\left(I_Y\right)^2}{s Y_h (\chi_{\pi\pi})^2}\,,\\
&\gamma_1=-\frac{4\pi I_Y\rho\left(sT+\mu\rho\right)}{s Y_h\left(\chi_{\pi\pi}\right)^2}-\mu\frac{\left(s T+k^2 I_Y\right)}{\left(\chi_{\pi\pi}\right)^2}Z_h\,,\\
&\Xi=\frac{4\pi \left(sT+\mu\rho\right)^2}{k^2 s Y_h\left(\chi_{\pi\pi}\right)^2}+\frac{\mu^2 Z_h}{\left(\chi_{\pi\pi}\right)^2}\,.
\end{split}
\end{equation}
We have verified that these expressions match the correlators computed numerically and that the phonons are dual to the vevs of the bulk fields $\psi_I$ \cite{Amoretti:2019}.
After turning on relaxation, these coefficients only receive small $O(\lambda)$ corrections, which can be safely neglected. 

\section{Relaxed WC hydrodynamics from holography}

To describe weak explicit breaking of translations, we assume $\lambda/\mu\ll\phi_{(v)}/\mu^2$. We also take $\lambda/\mu\ll k/\mu\ll1$. In our numerics, we choose $k/\mu=0.1$ and $\lambda/\mu=-10^{-5}$. 
{At zero wavevector $q$, WC hydrodynamics predicts  for the ac conductivity \cite{Delacretaz:2017zxd}} (see appendix \ref{app:WCreview})
\begin{equation}
\label{hydrosigma}
{\sigma(\omega)=\sigma_o+\frac{\omega_o^2\gamma_1(2\rho-i\gamma_1\chi_{\pi\pi}\omega)-\frac{\rho^2}{\chi_{\pi\pi}}(\Omega-i\omega)}{\omega^2-\omega_o^2+i\omega\Omega}\,.}
\end{equation}
We have set $\Gamma=0$, consistent with our numerics (see appendix \ref{app:kandlscaling}).
There are two gapped poles, which capture the relaxation of the phonons and of momentum: 
\begin{equation}
\label{omegapm}
\omega_\pm=-\frac{i}2\Omega\pm\frac12\sqrt{4\omega_o^2-\Omega^2}\,.
\end{equation}

\begin{figure}
\begin{tabular}{cc}
\includegraphics[width=.75\textwidth]{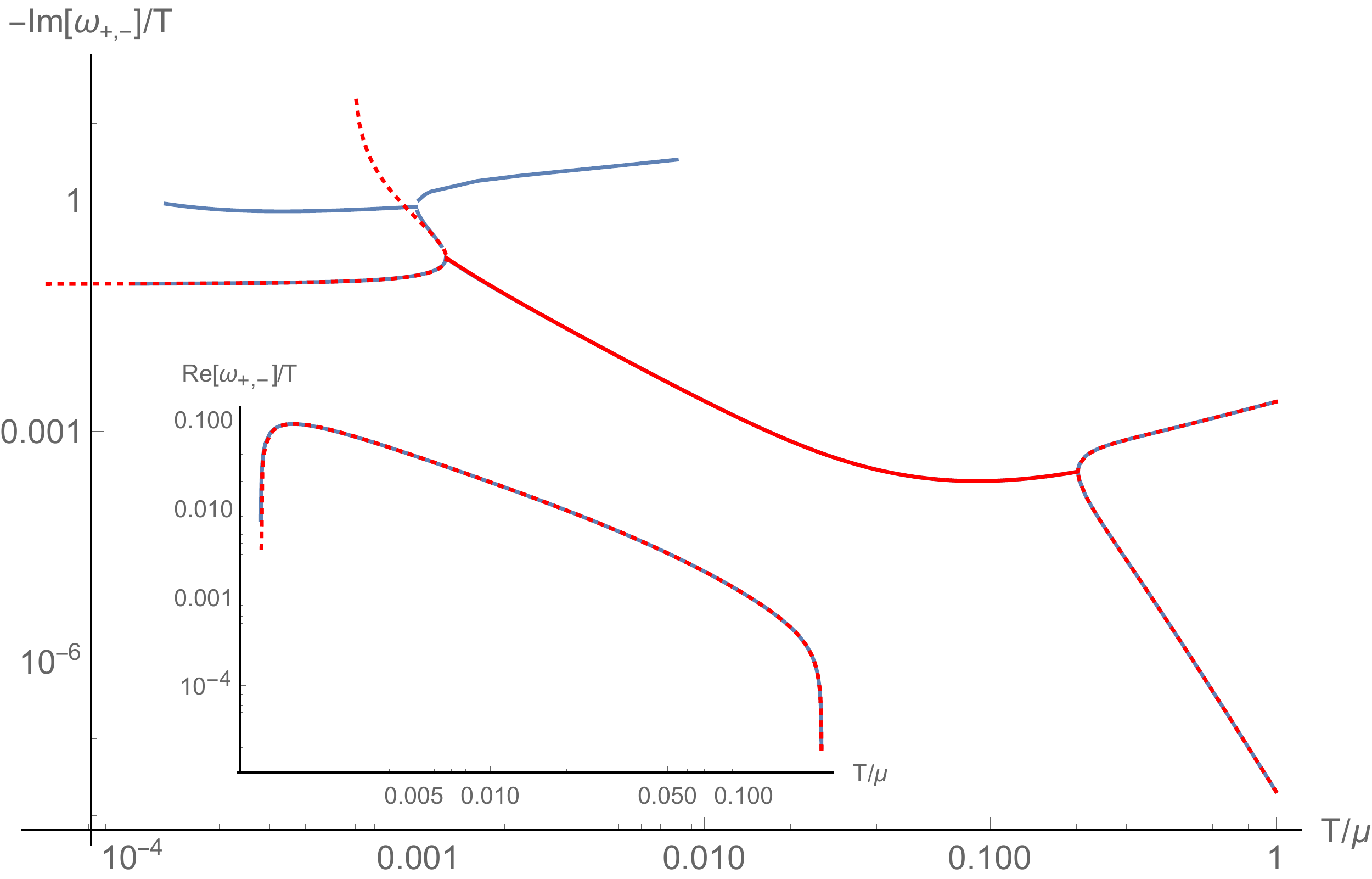}
\end{tabular}
\caption{$-\textrm{Im}[\omega_{\pm}]$ vs temperature. No visible difference between the exact numerics (solid, blue) and the analytical approximation using \eqref{omegapm} and \eqref{OmegaAnalytical} (dashed red).  Inset: real part.}
\label{fig:qnmvsT}
\end{figure}

In figure \ref{fig:qnmvsT}, we compute numerically (see appendix \ref{app:acqnm} for details) the lightest pair of quasi-normal modes (QNMs) of the holographic system  at zero wavevector and match their location to \eqref{omegapm}, see figure \ref{fig:OmegaAndomega0}.

{We also find that $\Omega\sim |\lambda|$, $\omega_o\sim|\lambda|^{1/2}k$ (see appendix \ref{app:kandlscaling}) and $G,K\sim k^2$, so that $m\sim|\lambda|^{1/2}$} \footnote{Similar Gell Mann-Oakes-Renner relations have previously been obtained in holography and in field theory \cite{Amoretti:2016bxs,Musso:2018wbv,Andrade:2018gqk}.}. This implies that both $\Omega$ and $m$ can be extracted in the limit $k=0$, ie small bulk and shear modulus. In this limit, the hydrodynamic correlator of the transverse phonon is
\begin{equation}
G^R_{\varphi\varphi}(\omega,q)=\frac{\Omega+q^2G\Xi}{G(q^2+ m^2)(i\omega-\Omega-q^2G\Xi)}\,.
\end{equation}
Near the boundary, $\delta\psi_{x}(r)= \delta\psi_{x,(0)}+\delta\psi_{x,(1)}r+O(r^2)$, where $\delta\psi_{x,(0)}$ is the source and $\langle O_{\delta\psi}\rangle=\lambda^2\delta\psi_{x,(1)}$ the vev. 
In appendix \ref{app:k=0}, we show that the phonon is $Gm^2\varphi=k\langle O_{\delta\psi}\rangle$. The pseudo-diffusive pole at $\omega\simeq-i\Omega -iG\Xi q^2$ is a consequence of the explicit breaking of the global shift symmetry when $\lambda\neq0$. 
We also compute:
\begin{equation}
\label{OmegaAnalytical}
\Omega\simeq\frac{I_Y m^2}{C_h Y_h}\simeq\frac1{\int_0^{r_h} dr\left(\frac{C_h Y_h \sqrt{B}}{C Y \sqrt{D}}-\frac1{4\pi T}\frac1{r_h-r}\right)}.
\end{equation}
The match to the exact numerics is excellent, see figures  \ref{fig:qnmvsT} and \ref{fig:OmegaAndomega0}.

\begin{figure}
\begin{tabular}{c}
\includegraphics[width=0.75\textwidth]{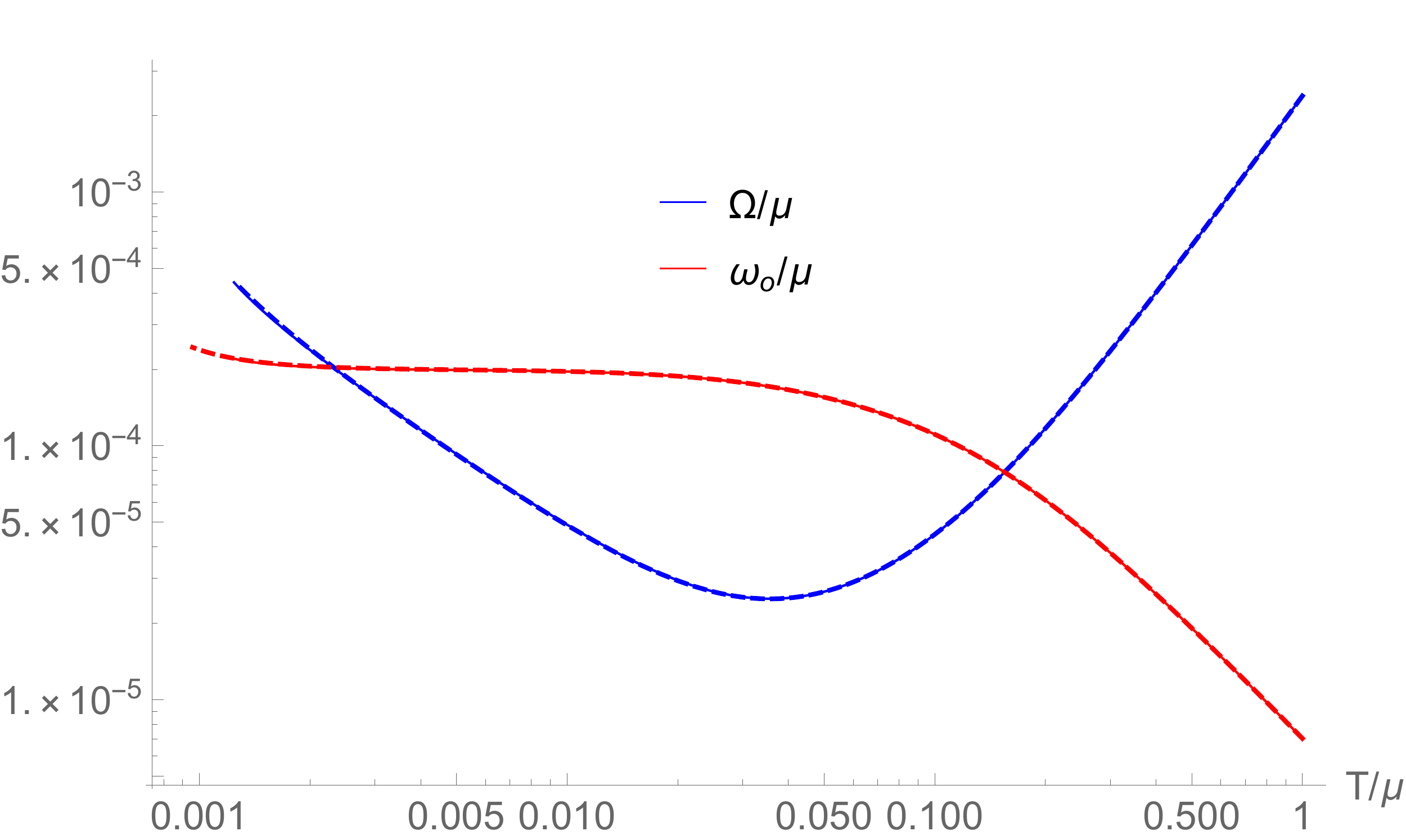}
\end{tabular}
\caption{$\Omega/\mu$ (blue) and $\omega_o/\mu$ (red) vs $T/\mu$. No visible difference between the exact numerical result from the QNMs location (solid line) and the analytical approximation \eqref{OmegaAnalytical} (dashed), evaluated on the $k\neq0$, $\lambda\neq0$ numerical background.}
\label{fig:OmegaAndomega0}
\end{figure}

At small $k/\mu$, \eqref{gamma1xiUnrelaxed} becomes $\Xi\simeq1/(k^2 C_h Y_h)+O(k^0)$. Putting this together with \eqref{OmegaAnalytical} leads to the relation \eqref{Omegarel} above.
This is one of our key results, and shows that the relaxation of the phonon is entirely governed by a diffusivity of the unrelaxed theory, in the entire range of temperatures where WC hydrodynamics applies. 

Assuming $\lambda\neq0$, the holographic Ward identity governing momentum relaxation is $\dot\pi^i=k\delta\psi_{i,(1)}$, which shows that $k$ controls the strength of momentum relaxation. Recall that $\omega_o^2\sim k^2$ while $\Omega\sim k^0$. {Restoring $\Gamma$ in the WC hydrodynamic expressions for the retarded Green's functions,}
\begin{equation}
\label{KuboPiDot}
\Gamma+\frac{\omega_o^2}{\Omega}=\frac{1}{\chi_{\pi\pi}}\lim_{\omega\to 0} \lim_{k\to 0}\frac1\omega\textrm{Im}G^R_{\dot\pi^x\dot\pi^x}(\omega,q=0)\,,
\end{equation}
\begin{equation}
    \frac{\omega_o^2}{\Omega^2}=-\frac{1}{\chi_{\pi\pi}}\lim_{\omega\to 0} \lim_{k\to 0}\textrm{Re} G^R_{\pi^x\pi^x}(\omega,q=0)\,.
\end{equation}
The Ward identity relates both quantities on the left hand side to $G^R_{\psi_x \psi_x}$ evaluated at $k=0$ (see appendix \ref{app:k=0}), leading to the same expressions as in \eqref{OmegaAnalytical}.

\section{Charge transport at high temperature}

As in many conventional systems where translations are broken spontaneously \cite{RevModPhys.60.1129}, our holographic system is an electrical insulator ($d\rho_{\textrm{dc}}/dT<0$) at high $T\gtrsim T_{\textrm{qc}}\simeq  5.10^{-2}\mu$.  

Its dc conductivity is $\sigma_{\textrm{dc}}=Z_h+\rho^2/(k^2 C_h Y_h)$, \cite{Donos:2014uba,Gouteraux:2014hca}.
Our numerics reveal that the dc conductivity is dominated by phase relaxation, $\sigma_{\textrm{dc}}\equiv\sigma(0)\simeq \rho^2\Omega/(m^2 G)$.

\begin{figure}
\begin{tabular}{cc}
\includegraphics[width=.75\textwidth]{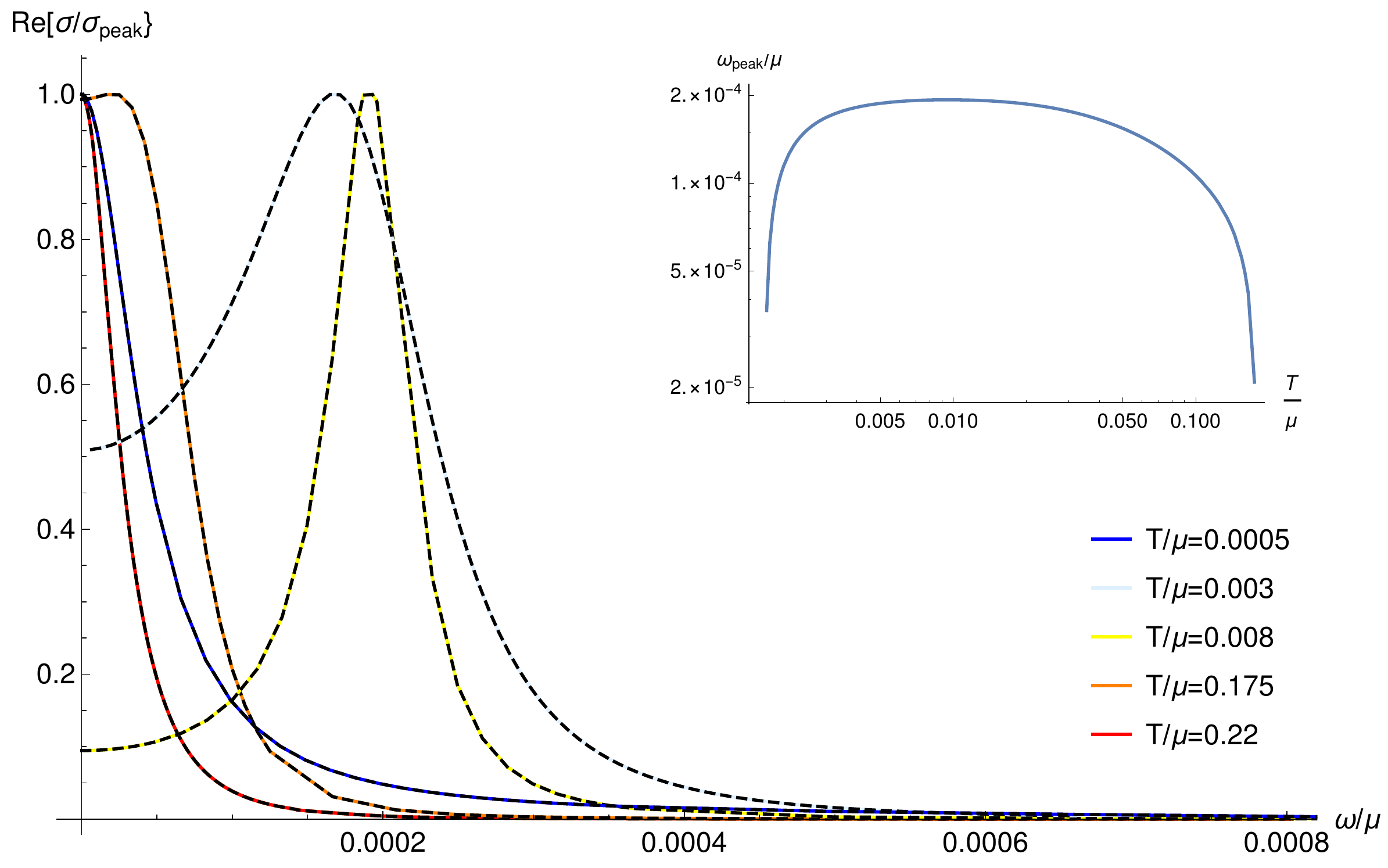}
\end{tabular}
\caption{No visible difference between $\textrm{Re}[\sigma]$ (normalized with respect to its maximal value $\sigma_{peak}$) computed numerically and hydrodynamic predictions \eqref{hydrosigma} and \eqref{OmegaXirel} (black dashed lines). The ac conductivity above $T_{\textrm{cm}}/\mu\simeq 10^{-3}$ shows an off-axis peak, which moves back on-axis as $T\lesssim T_{\textrm{cm}}$. Inset: temperature dependence of the peak location.}
\label{fig:conductivities}
\end{figure}

As $T$ decreases, the ac conductivity (see figure \ref{fig:conductivities}) shows a pinning peak moving away from zero frequency, $d\omega_{peak}/dT<0$. Correspondingly, the two poles $\omega_\pm$ collide, acquire a real gap and move away from the imaginary axis at low temperatures, see figure \ref{fig:qnmvsT}. The match between the lineshape of the ac conductivity and the hydrodynamic prediction \eqref{hydrosigma} is excellent.
$\Omega$ and $\omega_o$ are obtained from the location of the QNMs. We emphasize that there is no further fitting: $\sigma_o$ and $\gamma_1$ are computed using \eqref{gamma1xiUnrelaxed}, $\rho$ and $\chi_{\pi\pi}$ directly from the $\lambda\neq0$ numerical background.

\section{Charge transport at low temperatures}

For temperatures $T\lesssim T_{\textrm{qc}}\simeq  5.10^{-2}\mu $, the bulk geometry undergoes a qualitative change and becomes conformal to AdS$_2\times$R$^2$ near the horizon \cite{Amoretti:2017frz,Amoretti:2017axe}, with $s\sim T$ and metallic behavior  $\rho_{\textrm{dc}}\sim T$ -- see eg \cite{Anantua:2012nj,Davison:2013jba} and appendix \ref{app:lowTads2}. 

As temperature is lowered, the two off-axis QNMs turn around and move back towards the imaginary axis, see figure \ref{fig:qnmvsT}.

Accordingly, the off-axis peak in $\textrm{Re}\sigma(\omega)$ reverses direction, $d\omega_{peak}/dT>0$, and moves back towards zero frequency, see figure \ref{fig:conductivities}. In this region, small deviations between the hydrodynamic formula \eqref{hydrosigma} and the exact ac conductivity computed numerically are observed, together with a small mismatch between the dc values of at most $\sim1\%$. The first reason is that a third QNM is coming closer. The second reason are small but finite departures $O(\omega_o^2/(\Omega T)\sim k^2)$  from the hydrodynamic formula \eqref{hydrosigma} at low temperatures, similar to those reported in \cite{Davison:2015bea}. 
 In contrast, at high $T$ these corrections $O(\omega_o^2/(\Omega T))$ decay like a negative power of $T$, similar to \cite{Davison:2014lua}.

Eventually, the two QNMs collide, and one of them moves back up the imaginary axis, dominating the dynamics at very low temperatures. The other QNM goes down the imaginary axis and collides at $T_{\textrm{cm}}\simeq 10^{-3}\mu$ with the third QNM mentioned above, which signals the breakdown of WC hydrodynamics. {$T_{\textrm{cm}}$ is controlled by the magnitude of the explicit breaking scale $\lambda$ (see figure \ref{fig:compTc} in appendix \ref{app:k=0}). Below $T_{\textrm{cm}}$, the system behaves like a fluid with slow momentum relaxation with a Drude-like conductivity \cite{Hartnoll:2016apf}}
\begin{equation}
\label{OmegaXirel}
{\sigma(\omega)\simeq\frac{\rho^2}{\chi_{\pi\pi}}\frac1{\Gamma-i\omega}\quad\Rightarrow\quad \rho^{\textrm{dc}}\simeq \frac{\chi_{\pi\pi}}{\rho^2}\Gamma\,,}
\end{equation}
see figure \ref{fig:conductivities}. The longest-lived QNM captures slowly-relaxing momentum  and controls the width of the Drude peak. Its location is predicted to be $\omega\simeq-i \Gamma=-i k^2C_h Y_h/\chi_{\pi\pi}$ (setting $\omega_o=0$ in the right hand side of \eqref{KuboPiDot} for purely explicit breaking). In this regime, the phonons have decoupled from momentum: the two sub-dominant QNMs are very well approximated by the two longest-lived QNMs of the $k=0$ $\delta\psi$ correlator (the phonons), and contribute negligibly to $\textrm{Re}\sigma(\omega)$.

To summarize, we can distinguish between a high temperature phase with a crossover from insulating to metallic behavior as $T\lesssim T_{\textrm{qc}}$, dominated by damped, pinned phonons as predicted by WC hydrodynamics \cite{Delacretaz:2017zxd}; and a low temperature metallic phase at $T\lesssim T_{\textrm{cm}}$, where the charge dynamics of the system is simply governed by slow momentum relaxation, and the phonons are decoupled.

\section{Universal low temperature relaxation}

\begin{figure}
\begin{tabular}{cc}
\includegraphics[width=.75\textwidth]{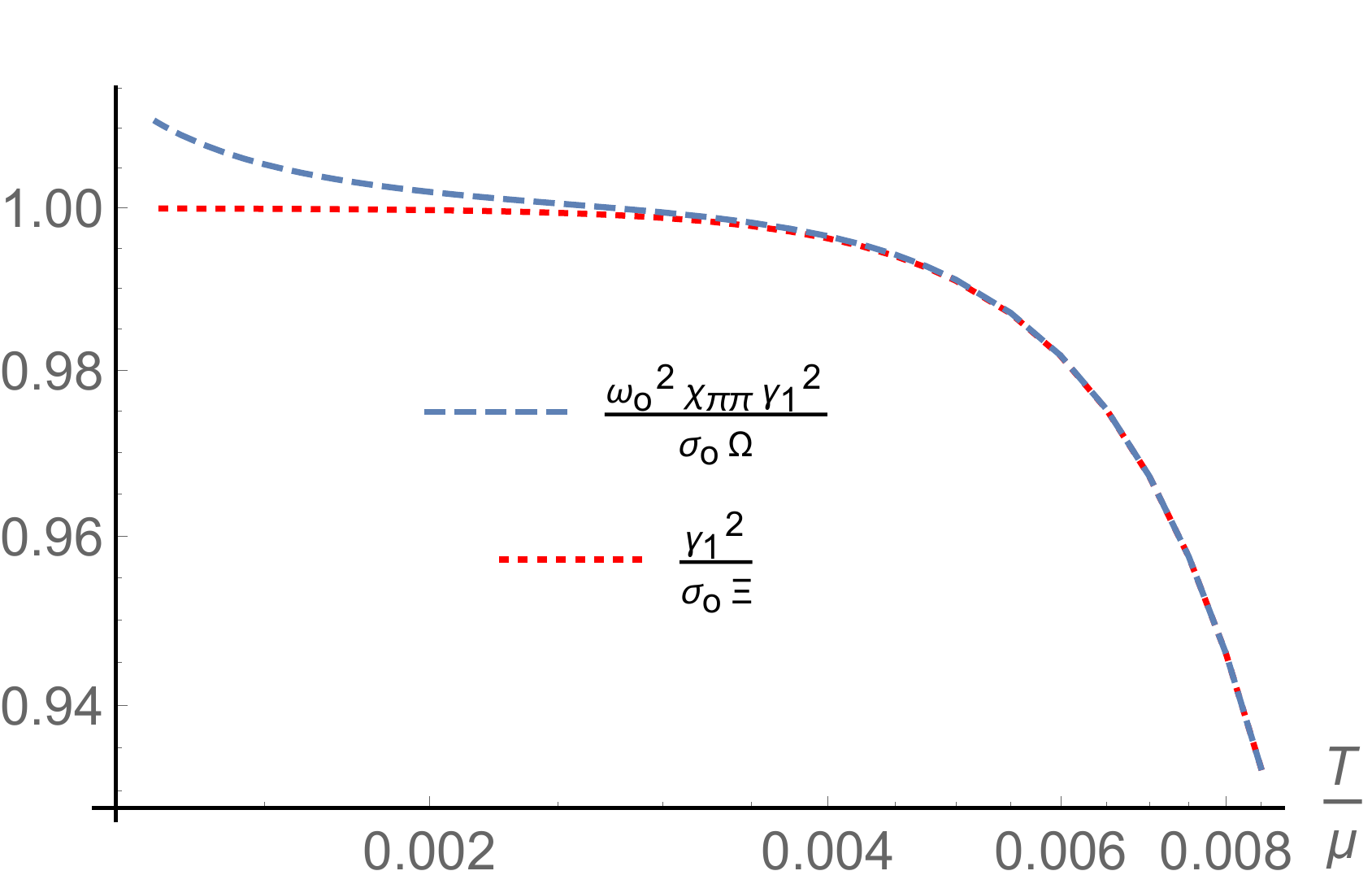}
\end{tabular}
\caption{The entropy bound of (un)relaxed WC hydrodynamics saturates as $T<T_{\textrm{qc}}$. As $T\to T_{\textrm{cm}}$, the bound of relaxed WC hydrodynamics starts to be violated.}
\label{fig:entropybound}
\end{figure}

Another universal aspect of relaxation in our model stems from positivity of entropy production in WC hydrodynamics, which places a bound on transport coefficients (see appendix \ref{app:WCreview})
\begin{equation}
\label{EntropyBound}
\gamma_1^2\leq \textrm{min}\left(\sigma_o\Xi,\frac{\Omega\sigma_o}{\chi_{\pi\pi}\omega_o^2}\right).
\end{equation}
Due to \eqref{Omegarel}, both terms on the right hand side are equal. At temperatures $T\lesssim T_{\textrm{qc}}$ marking the onset of the quantum critical phase, we observe from \eqref{gamma1xiUnrelaxed} (neglecting eg $sT$ terms vs $\mu\rho$ or $k^2 I_Y$) and our numerics (see figure \ref{fig:entropybound}) that
\begin{equation}
\label{lowTdiff}
\gamma_1\simeq-\frac{\mu}{\chi_{\pi j_q}}\sigma_o\,,\quad \Xi\simeq\frac{\Omega}{m^2 G}\simeq\left(\frac{\mu}{\chi_{\pi j_q}}\right)^2\sigma_o\,,
\end{equation}
where $\chi_{\pi j_q}=\chi_{\pi\pi}-\mu\rho$ since $j_q=\pi-\mu j$ (from relativistic symmetry) and $\chi_{j\pi}=\rho$. The expressions \eqref{lowTdiff} saturate the bound \eqref{EntropyBound}.

These results can be explained by considering the relaxation of the phonons into the heat current $j_q$, which is a universal mechanism in finite temperature systems. In particular, as translations are explicitly broken in our system, the only collective modes at the longest distances are diffusion of charge and energy. From WC hydrodynamics \cite{Delacretaz:2017zxd},
\begin{equation}
\label{xiKubo}
\Xi=\lim_{\omega\to0,q\to0,\lambda\to0}\frac{\textrm{Im}G^R_{\dot\varphi\dot\varphi}(\omega,q)}{\omega}\,,
\end{equation}
\begin{equation}
\label{OmegaKubo}
 \frac{\Omega}{\omega_o^2}=\chi_{\pi\pi}\lim_{\omega\to0,\lambda\to0}\frac{\textrm{Im}G^R_{\dot\varphi\dot\varphi}(\omega,q=0)}{\omega}\,.
\end{equation}
The limits $q\to0$ and zero relaxation $\lambda\to0$ do not commute. The $\omega\to0$ limit must be taken last. The universal contribution to the Hamiltonian $\Delta \mathcal H_q = \int d^2x (\pi\cdot j_q)/(\chi_{\pi j_q})$ gives $\dot\varphi=i[\Delta\mathcal H_q,\varphi]=j_q/\chi_{\pi j_q}$. This leads to \eqref{lowTdiff} using \eqref{xiKubo} and \eqref{OmegaKubo}: in this regime relaxation is controlled by thermal/incoherent diffusion processes \cite{Davison:2015taa,Davison:2018ofp,Davison:2018nxm}.

The small violation of the bound \eqref{EntropyBound} in figure \ref{fig:entropybound} as $T\to T_{\textrm{cm}}$ signals the breakdown of WC hydrodynamics. { As $\lambda\to0$, $T_{\textrm{cm}}\to0$ as well, so that in the spontaneous limit WC hydrodynamics applies to arbitrarily low temperatures and the bound $\gamma_1^2\leq \sigma_o\Xi$ saturates without violation \cite{Amoretti:2019}. In \cite{Amoretti:2019}, we also discuss how this depends on the scaling properties of the IR critical phase.}

\section{Outlook}

Experimental data in bad metals suggests $\omega_{peak}\sim k_B T/\hbar$, \cite{Delacretaz:2016ivq} ({see also \cite{2018PhRvB..98d5107W,PhysRevLett.120.087001} for more recent experimental observations)}.
In our holographic model, while $d\omega_{peak}/dT>0$, we do not observe $\omega_{peak}\sim k_B T/\hbar$. This is because in the region where $d\omega_{peak}/dT>0$, the phonon mass has a very weak dependence on temperature, while $\Omega\sim1/T$, consistent with the resistivity $\rho_{\textrm{dc}}\simeq m^2/\Omega\sim T$.

A linear temperature dependence $\omega_{peak}\sim T$ would obtain if $m\sim\Omega\sim T$ \cite{Delacretaz:2016ivq}, as expected for quantum fluctuations of an order parameter in the vicinity of a quantum critical point \cite{sachdev_2011}, provided $G\Xi\sim1/T$. For small bulk and shear modulus, this is indeed the behavior predicted by WC hydrodynamics in our system for the thermal diffusivity, $D_T\simeq\rho^2\Omega/(Gm^2)+O(G^0,K^0)\sim1/T$. In strongly-coupled systems, diffusivities are expected to saturate a lower bound $G\Xi\gtrsim \hbar v^2/k_B T$ \cite{Hartnoll:2014lpa}, where $v$ is some characteristic velocity. This is the behavior reported in recent experiments on cuprates and cold atoms \cite{2017PNAS..114.5378Z,2018arXiv180807564Z,2018arXiv180209456B}. It would be interesting to further investigate if holographic models of pinned density waves can capture this behavior.

Two distinct universal relaxation mechanisms are at play in our model. First, the smallness of the shear and bulk moduli implies $\Omega\simeq Gm^2\Xi$ in the whole temperature regime where WC hydrodynamics applies $T\gtrsim T_{\textrm{cm}}$. Independently, at low temperatures $T<T_{\textrm{qc}}$, the phonons relax into the heat current leading to \eqref{lowTdiff} (see also \cite{Delacretaz:2019}). Universal relaxation by hydrodynamic operators was discussed previously in \cite{Davison:2013txa,Lucas:2015lna} (see also \cite{Zaanen:2018edk}). This motivates a better characterization of the parameter space where these relaxation mechanisms dominate, including in other holographic models of pinned translational order  \cite{Jokela:2017ltu,Andrade:2017cnc,Alberte:2017cch,Andrade:2017ghg,Andrade:2018gqk,Donos:2019tmo}, in order to consider further their applicability to real materials.

\vskip1cm

\begin{acknowledgments}
We would like to thank Riccardo Argurio, Marco Fazzi, Saso Grozdanov, Javier Mas, Alfonso Ramallo and Javier Tarr\'\i o for stimulating and insightful discussions. We thank Richard Davison for comments on a previous version of the manuscript. We would also like to thank Tomas Andrade and Alexander Krikun for sharing with us an advanced draft of their work \cite{Andrade:2018gqk}, which has some overlap with our results. BG would especially like to thank Luca Delacr\'etaz, Sean Hartnoll and Anna Karlsson for numerous insightful discussions on Wigner crystal hydrodynamics and the various relaxation mechanisms for the phonons. BG has been partially supported during this work by the Marie Curie International Outgoing Fellowship nr 624054 within the 7th European Community Framework Programme FP7/2007-2013 and by the European Research Council (ERC) under the European Union’s Horizon 2020 research and innovation programme (grant agreements No 341222 and No 758759).  DM is funded by the Spanish grants FPA2014-52218-P and FPA2017-84436-P by Xunta de Galicia (GRC2013-024), by FEDER and by the Mar\'ia de Maeztu Unit of Excellence MDM-2016-0692. D.A. is supported by the `Atracci\'on del Talento' programme (Comunidad de Madrid) under grant 2017-T1/TIC-5258 and by Severo Ochoa Programme grant SEV-2016-0597 and FPA2015-65480-P (MINECO/FEDER). D.A. and D.M. thank the FRont Of pro-Galician Scientists for unconditional support. 
\end{acknowledgments}

\appendix

\section{Wigner crystal hydrodynamics\label{app:WCreview}}

We briefly recap the main features of Wigner crystal hydrodynamics, in the presence of weak explicit breaking of translations. More details can be found in \cite{chaikin_lubensky_1995,Delacretaz:2017zxd}.
As translations are broken spontaneously along both spatial directions, the usual conserved densities (energy, charge, momentum) need to be coupled to two Goldstone modes, $\varphi_i$, $i=x,y$. The free energy is supplemented by terms capturing the effect of the Goldstones:
\begin{equation}
f=\frac12 K|q\cdot\varphi(q)|^2+\frac12G\left(q^2+m^2\right)|\varphi(q)|^2=\frac12(K+G)\left|\lambda_\parallel(q)\right|^2+\frac{G}{2}\left|\lambda_\perp(q)\right|^2\,.
\end{equation}
$K$ and $G$ are the bulk and shear moduli, and characterize the stiffness of phase fluctuations around the ordered state. $m$ is the Goldstone mass generated by the explicit breaking of translations. It is convenient to parameterize the Goldstones by their longitudinal and transverse contributions, $\lambda_{\parallel}=\nabla\cdot\varphi$ and $\lambda_\perp=\nabla\times\varphi$. The corresponding sources $s_{\parallel,\perp}$ are defined by requiring that $\lambda_{\parallel,\perp}=\delta f/\delta s_{\parallel,\perp}$.

To leading order in gradients and keeping only linear terms, the Goldstones obey the following `Josephson' relations
\begin{equation}
\label{Josephson}
\begin{split}
\partial_t \lambda_\parallel+\Omega_\parallel\lambda_\parallel=&\nabla\cdot v+\gamma_1 \nabla^2\mu+\gamma_2 \nabla^2 T+\frac{\xi_\parallel}{K+G}\nabla^2 s_\parallel+\dots\,,\\
\partial_t \lambda_\perp+\Omega_\perp\lambda_\perp=&\nabla\times v+\frac{\xi_\perp}{G}\nabla^2 s_\perp+\dots\,,
\end{split}
\end{equation}
where $v$ is the velocity, $\mu$ the chemical potential, and $\gamma_{1,2}$ and $\xi_{\parallel,\perp}$ are diffusive transport coefficients. We have introduced two Goldstone damping rates $\Omega_\perp$ and $\Omega_\parallel$. They can in principle be distinct. For instance, in the presence of dislocations, the climb motion of dislocations, $\Omega_\parallel$, is suppressed \cite{Beekman:2016szb}.

These Josephson relations are supplemented by current, heat and momentum conservation equations (energy can be traded for entropy to linear order):
\begin{equation}
\label{Conseqhydro}
\partial_t \rho+\nabla\cdot j=0\,,\quad \partial_t s+\nabla\cdot (j_q/T)=0\,,\quad\partial_t\pi^i+\nabla_j T^{ji}=-\Gamma\pi^i-Gm^2\phi_i\,,
\end{equation}
together with the constitutive relations
\begin{equation}
\begin{split}
j=&\rho v-\sigma_o\nabla\mu-\alpha_o\nabla T-\gamma_1\nabla s_\parallel+\dots\,,\\
j_q/T=&s v-\alpha_o\nabla\mu-(\bar\kappa_o/T)\nabla T-\gamma_2\nabla s_\parallel+\dots\,,\\
T^{ij}=&\delta^{ij}\left(p+(K+G)\nabla\cdot\varphi\right)+2G\left[\nabla^{(i}\varphi^{j)}-\delta^{ij}\nabla\cdot\varphi\right]-\eta\left(2\nabla^{(i} v^{j)}-\delta^{ij}\nabla\cdot v\right)+\dots\,.
\end{split}
\end{equation}
The underlying conformal symmetry of the holographic setup implies that the stress-energy tensor is traceless, which sets the bulk viscosity to zero.

The hydrodynamic retarded Green's functions at nonzero frequency and wavevector are derived by following the Kadanoff-Martin procedure \cite{1963AnPhy..24..419K,Kovtun:2012rj}:
\begin{equation}
\label{GABKM}
G^R_{AB}(\omega,q)=M_{AC}\left[\left(i\omega-M\right)^{-1}\right]_{CD}\chi_{DB}
\end{equation}
where the vevs $A,B=(\delta\rho,\delta s,\pi_{\parallel},\lambda_\parallel,\pi_\perp,\lambda_\perp)$ and the corresponding sources are $(\delta\mu,\delta T,v_{\parallel}, s_\parallel,v_\perp, s_\perp)$. $M$ is the matrix
\begin{equation}\label{M}
M_{AB}=\left(\begin{array}{cccccc}
\sigma_o q^2&\alpha_o q^2&i \rho q&\gamma_1 q^2&0&0\\
\alpha_o q^2&\frac{\bar\kappa_o}{T}q^2 & i s q & \gamma_2 q^2 & 0 & 0 \\
i \rho q & is q&\eta q^2+\chi_{\pi\pi}\Gamma&iq&0&0\\
\gamma_1 q^2&\gamma_2 q^2&i q&\chi_{\lambda_\parallel\lambda_\parallel}\left(q^2\xi_\parallel+\Omega_\parallel\right)&0&0\\
0&0&0&0&\eta q^2+\chi_{\pi\pi}\Gamma&i q\\
0&0&0&0&iq&\chi_{\lambda_\perp\lambda_\perp}\left(q^2\xi_\perp+\Omega_\perp\right)
\end{array}\right).
\end{equation}
Relativistic symmetry of the holographic setup enforces that the momentum and energy current densities must be equal $\pi=j_e$, which places constraints on the transport coefficients:
\begin{equation}
\label{RelHydro}
\alpha_o=-\frac\mu{T}\sigma_o\,,\quad \bar\kappa_o=\frac{\mu^2}{T}\sigma_o\,,\quad \gamma_2=-\frac{\mu}T\gamma_1\,.
\end{equation}
Observe that this also means that the heat current $j_q\equiv j_e-\mu j=\pi-\mu j$.
Finally, the susceptibility matrix is\footnote{In general, nonzero static susceptibilities $\chi_{\rho\lambda_\parallel}$, $\chi_{s\lambda_\parallel}$ are also allowed. They play no role in our analysis, so we have set them to zero.}
\begin{equation}
\chi_{AB}=\left(\begin{array}{cccccc}
\chi_{\rho\rho}&\chi_{\rho s}&0&0&0&0\\
\chi_{\rho s}&\chi_{ss} & 0 &0 & 0 & 0 \\
0 & 0&\chi_{\pi\pi}&0&0&0\\
0&0&0&\frac{q^2}{K q^2+G (q^2+m^2)}&0&0\\
0&0&0&0&\chi_{\pi\pi}&0\\
0&0&0&0&0&\frac{q^2}{G(q^2+m^2)}
\end{array}\right)\,.
\end{equation}
A nonzero Goldstone mass implies that the Goldstone static susceptibilities $\chi_{\varphi_i\varphi_j}$ are finite in the limit $q\to0$ (rather than divergent like $1/q^2$ if $m=0$): there is no long range translational order in the system in the presence of explicit breaking.

Positivity of entropy production can be ensured by requiring all the eigenvalues of the matrix $M$ to be positive \cite{Delacretaz:2017zxd}. This implies: 
\begin{equation}
\eta\geq0,\qquad\sigma_o\geq0\,,\qquad \gamma_1^2\leq \sigma_o\frac{\xi_\parallel}{K+G}\,.
\end{equation}
Using \eqref{GABKM} and identities between the Green's functions stemming from \eqref{Conseqhydro} gives the retarded Green's functions quoted in equation \eqref{gamma1xiUnrelaxed} of the main text (with relaxation parameters turned off).
We defined {\it eg}
\begin{equation}
G^R_{\varphi_i \varphi_j}=\frac{q_i q_j}{q^4}G^R_{\lambda_\parallel\lambda_\parallel}+\left[\delta_{ij}-\frac{q_i q_j}{q^2}\right]\frac{G^R_{\lambda_\perp\lambda_\perp}}{q^2}\,.
\end{equation}
In this $q=0$ limit, and in absence of relaxation, for an isotropic crystal, $G^R_{\lambda_\parallel\lambda_\parallel}=G^R_{\lambda_\perp\lambda_\perp}$, 
since there should be no distinction between the longitudinal and transverse phonons. This leads to the constraint
\begin{equation}
\label{Xdef}
\frac{\xi_\parallel}{K+G}=\frac{\xi_\perp}{G}\equiv \Xi\,.
\end{equation}
Of direct interest to us is the presence in the longitudinal part of the spectrum of a diffusive mode, which in the limit of small bulk and shear moduli takes the simple expression:
\begin{equation}
\omega=-i\xi_{\parallel}q^2+O(K,G)+O(q^4)\,.
\end{equation}
This diffusive mode can be thought of as encoding the dissipation of the longitudinal Goldstone mode at long distances.

When translations are broken explicitly but the breaking is weak, the two Goldstone damping rates $\Omega_\parallel$, $\Omega_\perp$ become $q$-dependent (see \cite{Delacretaz:2017zxd}). In the long distance limit applicable to conductivities, $q\ll m$, they are equal $\Omega_\parallel=\Omega_\perp\equiv\Omega$.  Although explicit translation symmetry breaking was not considered as a microscopic origin for $\Omega$ in \cite{Delacretaz:2017zxd}, the functional form of the hydrodynamic equations and retarded Green's functions is insensitive to the microscopic origin of such terms. Phonon damping by disorder was studied very early on, see eg \cite{PhysRevB.17.535,PhysRevB.28.340}, but was reported to affect the ac conductivity only through a pinning mass $m$ and momentum relaxation rate $\Gamma$ \cite{RevModPhys.60.1129}. Here we show that it also leads to phase relaxation $\Omega$ in the sense of \cite{Delacretaz:2016ivq,Delacretaz:2017zxd} and a nonzero dc conductivity. This is well-known from studies of the magnetic field-induced melting of Wigner solids \cite{PhysRevB.18.6245,PhysRevB.62.7553}

The conductivity equation \eqref{hydrosigma} quoted in the main text is defined from the retarded Green's function of the current
\begin{equation}
\sigma(\omega)\equiv\frac{i}{\omega}G^R_{jj}(\omega,q=0)\,.
\end{equation}
The poles in the denominator of this expression are given in equation \eqref{omegapm} of the main text, with the pinning frequency defined as $\omega_o^2=Gm^2/\chi_{\pi\pi}$ and setting $\Gamma=0$. For reference, we quote here its full expression without neglecting terms proportional to the momentum relaxation rate $\Gamma$:
\begin{equation}\label{hydrocondu}
\sigma(\omega)=\sigma_o+\frac{\frac{\rho^2}{\chi_{\pi\pi}}(\Omega-i\omega)-\omega_o^2\gamma_1\left[2\rho+\gamma_1\chi_{\pi\pi}\left(\Gamma-i\omega\right)\right]}{(\Gamma-i\omega)(\Omega-i\omega)+\omega_o^2}\,.
\end{equation}
The Drude formula in equation \eqref{OmegaXirel} of the main text is recovered by setting the pinning frequency to zero or equivalently sending $\Omega\to+\infty$.

\section{Numerical methods \label{app:NumMeth}}

We have computed numerically black hole solutions to the action 
\begin{equation}
S=\int d^{4}x\,\sqrt{-g}\left[R-\frac12\partial\phi^2-V(\phi)-\frac14 Z(\phi)F^2-\frac12\sum_{I=1}^{2}Y(\phi)\partial \psi_I^2\right]\,.
\end{equation}
We adopt the following Ansatz for the metric and matter fields 
\begin{align}
&ds^2={1\over r^2}\left(-u(r)dt^2+{1\over u(r)}dr^2+c(r)(dx^2+dy^2)\right)\,,
\label{eq:metric}\\
&A=A_t(r)dt\,,\quad \phi=\phi(r)\,,\quad \psi_I=k x^I\,,\quad x^I=\{x,y\}\,,
\end{align}
which allows for solutions that break translations pseudo-spontaneously.
This follows from the Ansatz for the scalars $\psi_I$ and the asymptotic behavior of the scalar coupling $Y(\phi)$, as explained in the main text.

The resulting equations of motion can be reduced to a system of four ordinary differential equations (three are second order and one is first order). The scalar couplings behave in the UV as
\begin{equation}
V_{uv}(\phi)=-6-\phi^2+O(\phi^3), \quad Z_{uv}(\phi)=1+O(\phi), \quad Y_{uv}(\phi)=\phi^2+O(\phi^3)\,.
\end{equation}
{The asymptotic behavior of the metric and matter fields is
\begin{subequations}
	\begin{align}
	&\phi(r)=\lambda\,r+v\,r^2+{1\over36c_0^2}\left[
	-9c_1\,(c_1\,\lambda+4c_0\,v)+c_0\,\lambda\,(36k^2+7c_0\,\lambda^2)
	\right]\,r^3+O(r^4)\,,\label{eq:phiuv}\\
	&A_t(r)=\mu-\rho\,r+O(r^2)\,,\\
	&u(r)=1+{c_1\over c_0}\,r+{1\over4}\left({c_1^2\over c_0^2}-\lambda^2\right)r^2+u_3\,r^3+O(r^4)\,,\\
	&c(r)=c_0+c_1\,r+{1\over4c_0}\,\left(c_1^2-c_0^2\,\lambda^2\right)r^2
	-{\lambda\over6}\left(c_1\,\lambda+2c_0\,v\right)r^3
	+O(r^4)\,,
	\end{align}
	\label{eq:uvback}
\end{subequations}
where higher order coefficients are functions of $\lambda$, $v$, $\rho$,
$u_3$, $c_0$, and $c_1$.
Notice that in order for this solution to asymptote to AdS we must have $c_0=1$, and
$c_1 = 0$.}
$\rho$ is the charge density, $\mu$ the chemical potential and $v$ is related to the vev of the scalar $\phi$.

We can also expand the solution near the black hole horizon $r=r_h$:
\begin{subequations}
	\begin{align}
	&\phi(r)=\phi_{h}+O(r_h-r)\,,\qquad\qquad
	A_t(r)=A_{h,1}(r_h-r)+O((r_h-r)^2)\,,\\
	&u(r)=u_{h,1}(r_h-r)+O((r_h-r)^3)\,,\quad
	c(r)=c_h+c_{h,1}(r_h-r)+O((r_h-r)^2)\,.
	\end{align}
	\label{eq:irback}
\end{subequations}
In our numerics, we choose for the potentials
\begin{equation}
 V(\phi)=-6\cosh(\phi/\sqrt3), \quad Z(\phi)=\exp(-\phi/\sqrt3), \quad Y(\phi)=(1-\exp\phi)^2\,.
\end{equation}
Then
\begin{equation}
u_{h,1}={c_h\,e^{-{\phi_h\over\sqrt{3}}}
	\left[6+e^{2\phi_h\over\sqrt{3}}(6-r_h^4\,A_{h,1})
	\right]
	-2k^2r_h^2\left(1-e^{\phi_h}\right)^2
	\over 2r_h\left(2c_h+c_{h,1}\,r_h\right)}
\end{equation}
determines the temperature of the black hole $T=-u_{h,1}/(4\pi)$, and further higher
order coefficients in \eqref{eq:irback} are also determined in terms of $\phi_h$,
$A_{h,1}$, $c_h$, and $c_{h,1}$.

By using the scale invariance
$(t,x,y,r)\to\alpha\,(t,x,y,r)$, $A_t\to A_t/\alpha$, $k\to k/\alpha$ of the equations of motion, we can set the horizon radius $r_h=1$ in our numerical computations.
Numerical solutions are generated by integrating the equations of motion from the IR ($r=1$) to the UV ($r=0$). Using another scale invariance of the equations under $(x,y)\to \beta\,(x,y)$, $k\to k/\beta$,
$c\to c/\beta^2$, we set $c_0 = 1$.
This fixes $c_h$, leaving three free IR parameters $\phi_h$, $A_{h,1}$, $c_{h,1}$; and one UV condition: $c_1=0$.
{Therefore, for each value of $k$
we expect to obtain a two-parameter family of solutions. We can choose those parameters to be the dimensionless ratios $T/\mu$, $\lambda/\mu$.
All in all we can parametrize the space of solutions in terms of the three 
dimensionless ratios $T/\mu$, $\lambda/\mu$, and $k/\mu$.}

In this work we are interested in solutions breaking translations pseudo-spontaneously.
These are geometries where $\lambda/\mu\ll v/\mu^2$.

\subsection{Low temperature, near horizon geometry \label{app:lowTads2}}
As we decrease the temperature below $T_{\textrm{qc}}\simeq 5.10^{-2}\mu$, the scalar $\phi$ diverges near the horizon. The scalar couplings become
\begin{equation}
\label{IRscalarcouplings}
V_{IR}=-3 e^{-\phi/\sqrt{3}}\ ,\quad Z_{IR}= e^{-\phi/\sqrt{3}}\ ,\quad Y_{IR}=1\ ,
\end{equation}
leading to a solution of the equations of motion conformal to AdS$_2\times R^2$, with a dynamical Lifshitz exponent $z=+\infty$ \cite{Anantua:2012nj,Davison:2013txa,Gouteraux:2014hca,Amoretti:2017frz,Amoretti:2017axe}:
\begin{equation}\label{skaska}
\begin{split}
& ds^2 = \frac1\zeta \left[ -\frac{f(\zeta)}{\zeta^2}L_t^2dt^2 + \frac{\tilde L^2 d\zeta^2}{\xi^2f(\zeta) } + L_x^2dx^2+L_x^2dy^2\right], \quad 
 A =a_o\, \zeta^{-2} dt\,,\\
& \phi=-\sqrt{3}\log \zeta\,,\quad f(\zeta)=1-\left(\frac{\zeta}{\zeta_h}\right)^{2}.
 \end{split}
\end{equation}
This solution is valid for values of the IR radial coordinate $\zeta\gg\mu$. The constants $L_t$, $L_x$, $\tilde L$, $a_o$ can be determined by matching to the UV AdS$_4$ asymptotics.
In these coordinates, $T\sim \zeta_h^{-1}$, so that the entropy density $s\sim \zeta_h^{-1}\sim T$.

\begin{figure}
\begin{tabular}{c}
\includegraphics[width=.75\textwidth]{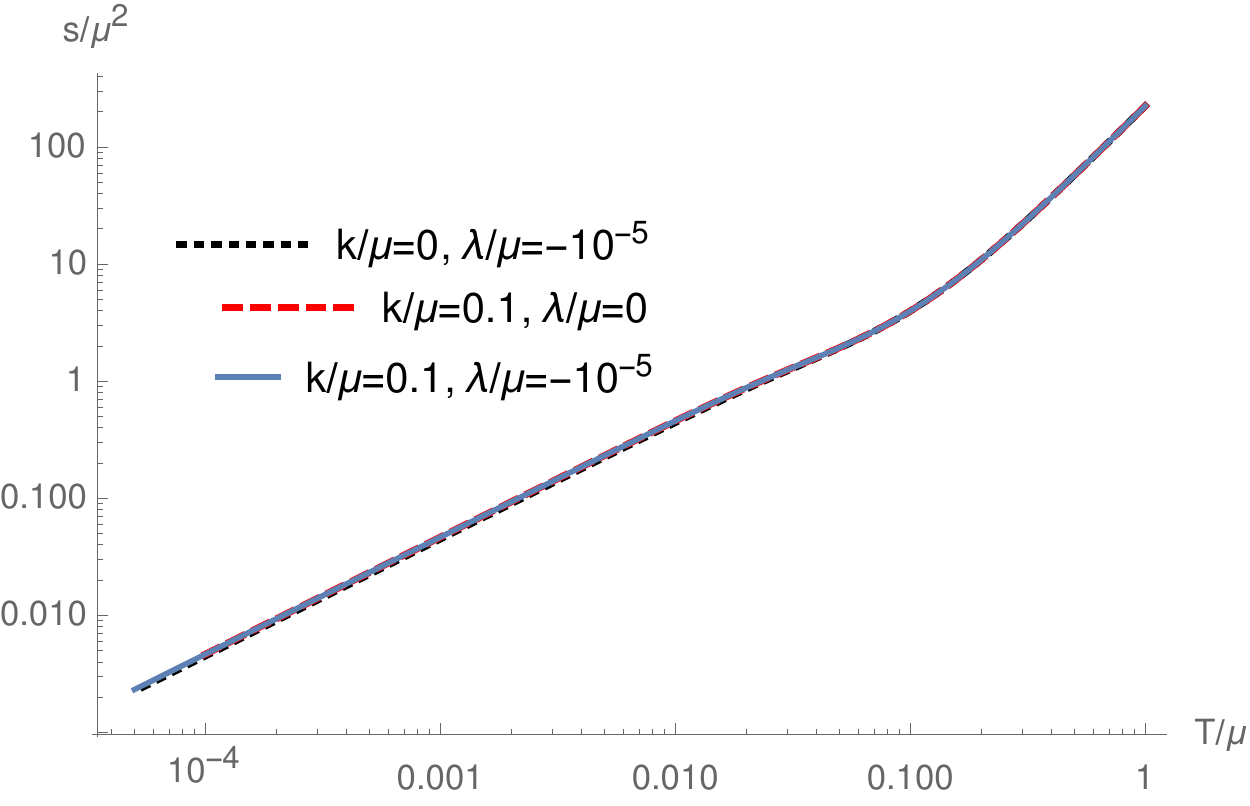}
\end{tabular}
\caption{{Comparison of the entropy density vs temperature for various values of $k$ and $\lambda$.}}
\label{fig:svsT}
\end{figure}

{A rough estimate of the temperature $T_{\textrm{qc}}$ where the near horizon geometry becomes conformal to AdS$_2\times R^2$ is given by the temperature below which the entropy density becomes linear in temperature. This is displayed in figure \ref{fig:svsT} for various values of $k$ and $\lambda$. All curves fall on top of one another, which demonstrates that $T_{\textrm{qc}}$ is relatively insensitive to the specific value of $k$ and $\lambda$ for the typical ranges we are interested in. }

\subsection{AC Correlators: Conductivity and QNMs \label{app:acqnm}}
In order to compute the conductivity of the boundary theory and the QNMs at zero momentum,
it is sufficient to study the following consistent set of fluctuations
\begin{align}
\delta g_{tx}=h(r)\,e^{-i\omega t}\,,\quad \delta A_x=a(r)\,e^{-i\omega t}\,,\quad
\delta \psi_x=
\xi(r)\,e^{-i\omega t}\, .
\label{eq:ACflucs}
\end{align}
To linear order in the fluctuations, the equations of motion
for $a$, $h$, and $\xi$ are a consistent set formed by two second order and one
first order differential equation.

For a pseudo-spontaneous background geometry with $\lambda\neq0$, the fluctuations have the following UV expansion
\begin{subequations}
	\begin{align}
	&h(r)=r^{-2}\left(h_0+O(r^3)\right)\,, \\ 
	&a(r)=a_0+a_1\,r+O(r^2)\,, \\
	&\xi(r)=\xi_0+\xi_1 r+ O(r)\,,
	\end{align}
	\label{eq:flucUV}
\end{subequations}
where higher order coefficients are functions of $h_0$, $a_0$, $a_1$, $\xi_{1}$
and $\xi_0$.

The AC conductivity can be read from $G^R_{j^xj^x}$ through a Kubo formula which
in terms of the asymptotic expansion \eqref{eq:flucUV} takes the form
\begin{equation}
\sigma(\omega)={1\over i\omega}\,G^R_{j^xj^x}(\omega)={a_1\over i\omega\,a_0}\,.
\label{eq:sigmaAC}
\end{equation}
The second equality holds for configurations where 
the only nonzero source 
is given by $a_0$ which corresponds to turning on an electric field along $x$.
The only other independent source in the UV solution \eqref{eq:flucUV} is given
by the diffeomorphism-invariant combination
\begin{equation}
\xi_0-{ik\over\omega}\,h_0\,,
\label{eq:flucUVsource}
\end{equation}
and it should vanish on the solutions used to compute $\sigma(\omega)$ through \eqref{eq:sigmaAC}.

We are interested
in computing a retarded correlator, hence we look for solutions with ingoing
boundary conditions at the black hole horizon. They read
\begin{subequations}
	\begin{align}
	&h(r)\,e^{i{\omega\over u_{h,1}}}=h_{h,1}\,(r_h-r)+O((r_h-r)^2)\,, \\ 
	&a(r)\,e^{i{\omega\over u_{h,1}}}=a_h+O(r_h-r)\,, \\
	&\xi(r)\,e^{i{\omega\over u_{h,1}}}=\xi_h+O(r_h-r)\,,
	\end{align}
	\label{eq:flucIR}
\end{subequations}
with
\begin{equation}
h_{h,1}={a_h\,A_{h,1}\,e^{-{\phi_h\over\sqrt{3}}}\,r_h^2-\left(
	e^{\phi_h}-1\right)^2 \xi_h\,k/r_h
	\over i\omega/u_{h,1}-1}\,,
\end{equation}
and other higher order coefficients determined as well in terms of $a_h$, and $\xi_h$.

We construct numerical solutions by integrating the
system of three linear differential equations for $h,a,\xi$ from the horizon ($r=r_h=1$)
to the boundary ($r=0$). There are two free IR parameters in \eqref{eq:flucIR}, but since
the equations are linear we can scale away one of them (eg set $a_h=1$). Therefore we are left with
one free (complex) parameter and one UV boundary condition: we shoot for solutions where
the combination \eqref{eq:flucUVsource} vanishes, and read $\sigma(\omega)$ using
\eqref{eq:sigmaAC}.

We shall next compute the QNMs given by the poles of the holographic correlator 
$G^R_{AB}(\omega)$ where $A,B =h,a,\xi$ are the set of fields \eqref{eq:ACflucs}.
We follow
\cite{Kaminski:2009dh} and employ
the so-called determinant method. Hence we need to obtain three independent solutions
for the fluctuations, and construct the following matrix of sources
\begin{equation}
S=\left(
\begin{array}{ccc}
h_0^{(I)} & h_0^{(II)}  & -i\omega \\
a_0^{(I)} & a_0^{(II)}  & 0\\
\xi^{(I)}_{-1} & \xi^{(II)}_{-1} & k
\end{array}
\right)\,,
\label{eq:sourcematrixtrans}
\end{equation}
where in order to generate the third column we have used the pure gauge solution
\begin{equation}
h(r)=-i\omega\,{c(r)\over r^2}\,,\quad
a(r)=0\,, \quad \xi(r)=k\,.
\label{eq:puregaugefluctrans}
\end{equation}
It is straightforward to construct two independent numerical solutions integrating
from the horizon with boundary conditions \eqref{eq:flucIR}.

Finally, the QNMs, namely the complex frequencies where the holographic Green functions have a pole, are given by the values of $\omega$ for which the determinant of the
matrix \eqref{eq:sourcematrixtrans} vanishes~\cite{Kaminski:2009dh}.

\section{$\psi$ correlators at $k=0$ \label{app:k=0}}

In this appendix, we solve the equations of motion and compute the correlators of the fields $\psi_I$ in the limit $k=0$. Similar results have also been reported in \cite{Donos:2019txg}. In this case, both fields decouple and so we will just denote them by $\delta\psi$. The equation of motion for the fluctuation $\delta\psi=\delta\psi(r) \exp(-i\omega t+i q x)$ is
\begin{equation}
\label{deltapsieq}
\left(\sqrt{\frac{D}{B}}CY \delta\psi'\right)'+\left(\omega^2-\frac{D}{C}q^2\right) C Y\sqrt{\frac{B}D}\delta\psi=0\,.
\end{equation}
Using a standard procedure in holography \cite{Policastro:2002se,Policastro:2002tn,Herzog:2003ke}, this equation can be solved perturbatively at small $\omega,q$ compared to $r_h$, assuming the presence of a regular black hole horizon. We consider the expansion of the perturbation
\begin{equation}
\label{deltapsiexp}
\delta\psi(r)=\left(1-\frac{r}{r_h}\right)^{-\frac{i\omega}{4\pi T}}\left(1+\frac{i\omega}{4\pi T}\delta\psi_1+\left(\frac{q}{4\pi T}\right)^2\delta\psi_2+O\left(q^3,\omega^3,\omega q^2\right)\right).
\end{equation}
After transforming to Eddington-Finkelstein coordinates, this corresponds to the ingoing mode at the horizon \cite{Son:2002sd}.
The horizon is located at $r=r_h$ and the boundary at $r=0$. Plugging \eqref{deltapsiexp} into \eqref{deltapsieq}, the equation is solved perturbatively order by order in $\omega$.

The order $\omega^0$ is trivially solved by a constant, which can always be set to unity by linearity of the equation of motion.
The solution at $O(\omega$) reads
\begin{equation}
\label{deltapsi1sol}
\delta \psi_1(r)=\int_{r_h}^r dr\left(\frac{4\pi T C_h Y_h \sqrt{B}}{C Y \sqrt{D}}-\frac1{r_h-r}\right),
\end{equation}
where we have imposed horizon regularity to fix one of the integration constants, and chose the other one such that $\delta \psi_1(r_h)=0$ (which otherwise would simply correspond to a different choice of normalization of the source). 

The solution for $\delta \psi_2$ after imposing horizon regularity is
\begin{equation}
\label{deltapsi2sol}
\delta\psi_2=(4\pi T)^2\int_{r_h}^r\frac{\sqrt B}{\sqrt D CY}\int_{\tilde r_h}^{\tilde r}\sqrt{BD}Y\,.
\end{equation}

\begin{figure}
\begin{tabular}{c}
\includegraphics[width=.75\textwidth]{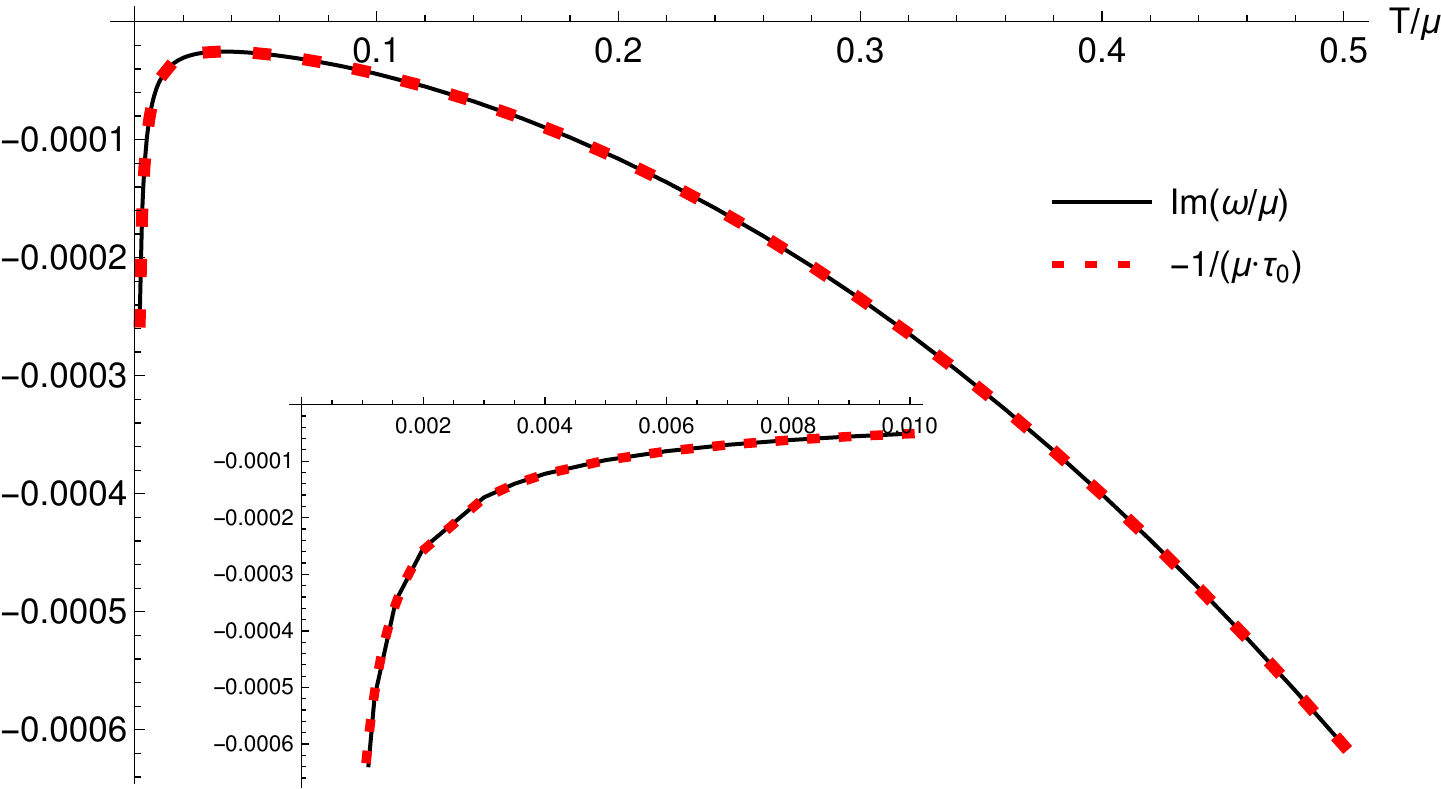}
\end{tabular}
\caption{Numerical check that the gap in \eqref{disprelpsik=0} at $q=0$ is well-captured by \eqref{tau0}. Data for $\lambda/\mu=-10^{-5}$.}
\label{fig:1ovTau}
\end{figure}

\begin{figure}
\begin{tabular}{c}
\includegraphics[width=.75\textwidth]{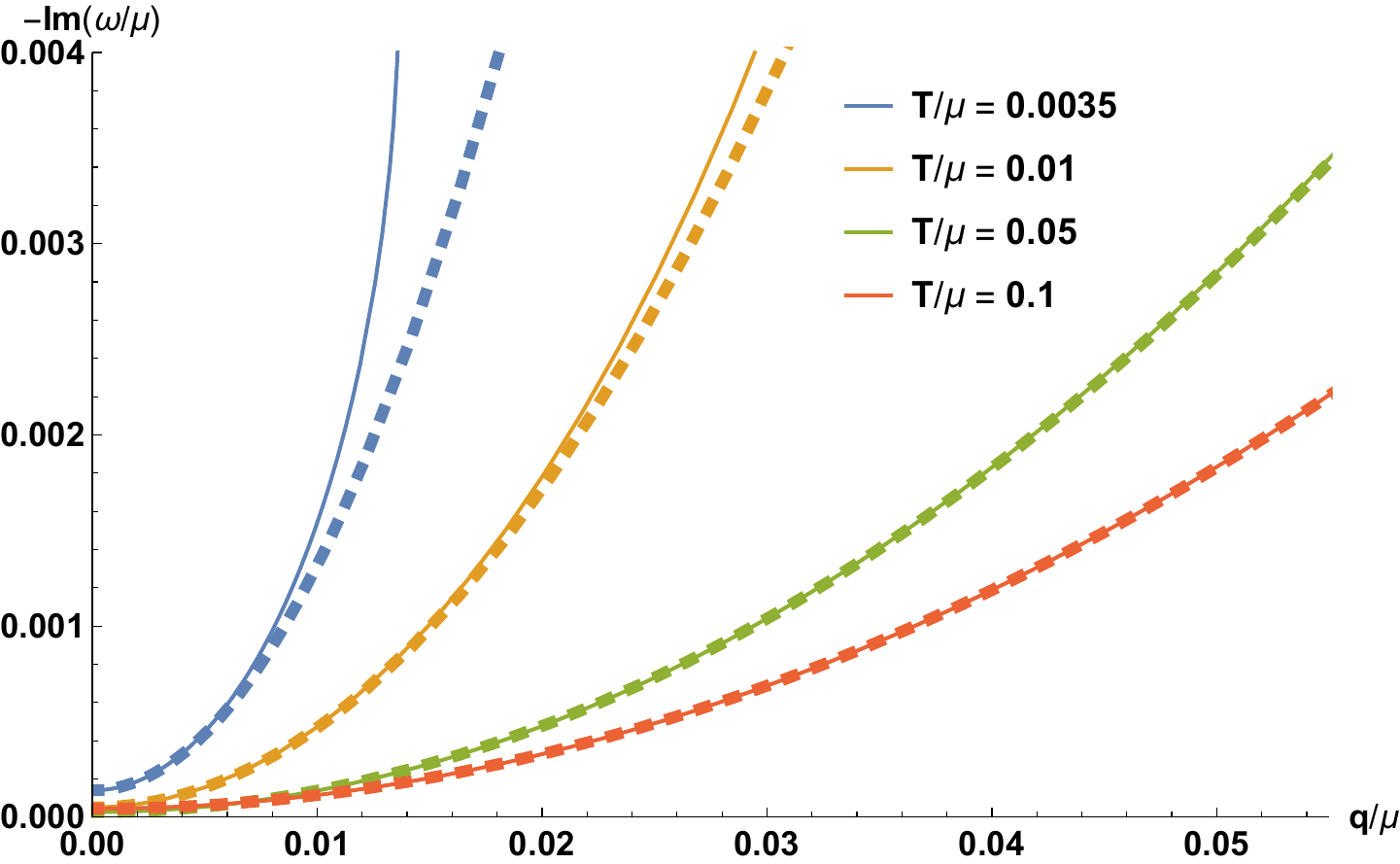}
\end{tabular}
\caption{Numerical check of the dispersion relation \eqref{disprelpsik=0}.
Solid lines represent the numerical
data and dashed lines correspond to equation \eqref{disprelpsik=0}. }
\label{fig:Disprelk0}
\end{figure}

We now put everything together back in \eqref{deltapsiexp} and expand close to the boundary. For nonzero source $\lambda\neq0$, the asymptotic expansion of $\delta\psi$ takes the form
\begin{equation}
\label{UVexppsilambda!=0}
\delta\psi(r\to0)=\delta\psi^{(0)}+r\delta\psi^{(1)}+\ldots
\end{equation}
where the $r^0$ term is the source and the $r^1$ term is related to the vev by
\begin{equation}
\langle O_{\delta\psi}\rangle = \lambda^2\delta\psi^{(1)}\,.
\end{equation}
We find
\begin{equation}
\label{GRpsipsilambdanon0q}
G^R_{\delta\psi\delta\psi}=\frac{i\omega C_h Y_h-q^2 I_Y}{1-i\omega\tau_o+\tau_o D_\psi q^2}\,,
\end{equation}
with
\begin{equation}
\label{tau0}
\tau_o=\int_0^{r_h} dr\left(\frac{C_h Y_h \sqrt{B}}{C Y \sqrt{D}}-\frac1{4\pi T}\frac1{r_h-r}\right),
\end{equation}
and
\begin{equation}
  D_\psi=\frac1{\tau_o}\int_{r_h}^0\frac{\sqrt B}{\sqrt D CY}\int_{r_h}^{\tilde r}\sqrt{BD}Y\,.
\end{equation}
This gives a pseudo-diffusive pole located at
\begin{equation}
\label{disprelpsik=0}
\omega_{k=0,\lambda\neq0}=
-i\Omega-iD_\psi q^2+O(q^4)\,,\quad \Omega=\frac1{\tau_o}\,,
\end{equation}
which matches very well the exact location of the pole determined numerically, see figures \ref{fig:1ovTau} and \ref{fig:Disprelk0}. The exact numerical dispersion relation deviates from \eqref{disprelpsik=0} as $q$ increases or as $T$ decreases, which is expected. Actually, at sufficiently low $T$ or large $q$, the pole collides with another purely imaginary pole and moves off axis. This collision and subsequent motion of the poles matches very well the motion of the two CM poles in figure 1 of the main text at the lowest temperatures $T<T_{\textrm{cm}}$. 

\begin{figure}
\begin{tabular}{c}
\includegraphics[width=.75\textwidth]{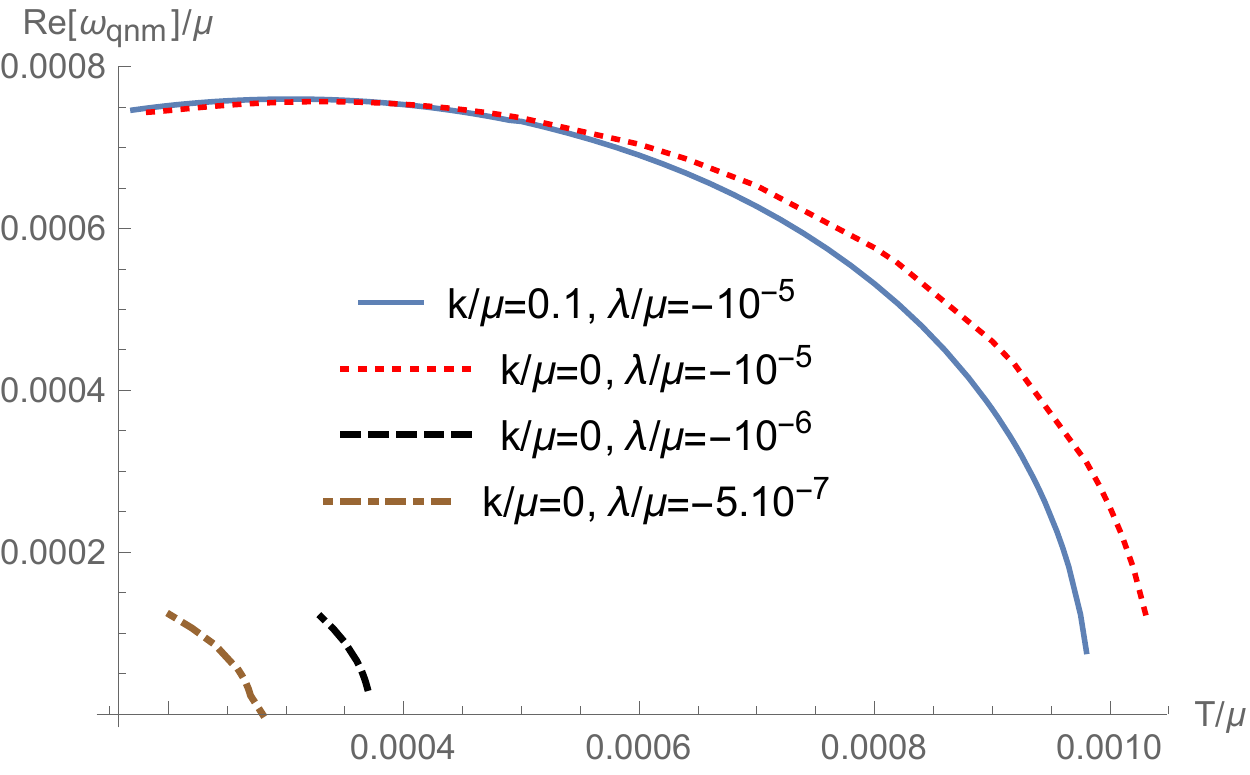}
\end{tabular}
\caption{{Real part of the longest-lived complex QNMs at low temperatures $T\leq T_{\textrm{cm}}$ ($ T_{\textrm{cm}}$ is the temperature at which the real part vanishes), showing the dependence of $T_{\textrm{cm}}$ on $k$ and $\lambda$.}}
\label{fig:compTc}
\end{figure}

{In figure \ref{fig:compTc}, we compare the collision temperature $T_{\textrm{cm}}$
at $k=0$ and $k/\mu=0.1$ by plotting the real part of the relevant QNMs close to the collision. At $\lambda/\mu=-10^{-5}$, there is little variation between $k=0$ and $k/\mu=0.1$. On the other hand, varying $\lambda/\mu$ from $-10^{-5}$ to $-5.10^{-7}$ shows that $T_{\textrm{cm}}$ depends in a much more pronounced way on the value of $\lambda$, which is expected since $\tau_o^{-1}\sim\lambda$ (see top row of figure \ref{fig:parvsl} below). Indeed, $T_{\textrm{cm}}=0$ in the purely spontaneous limit $\lambda=0$, \cite{Amoretti:2019}. }

\begin{figure}
\begin{tabular}{c}
\includegraphics[width=.75\textwidth]{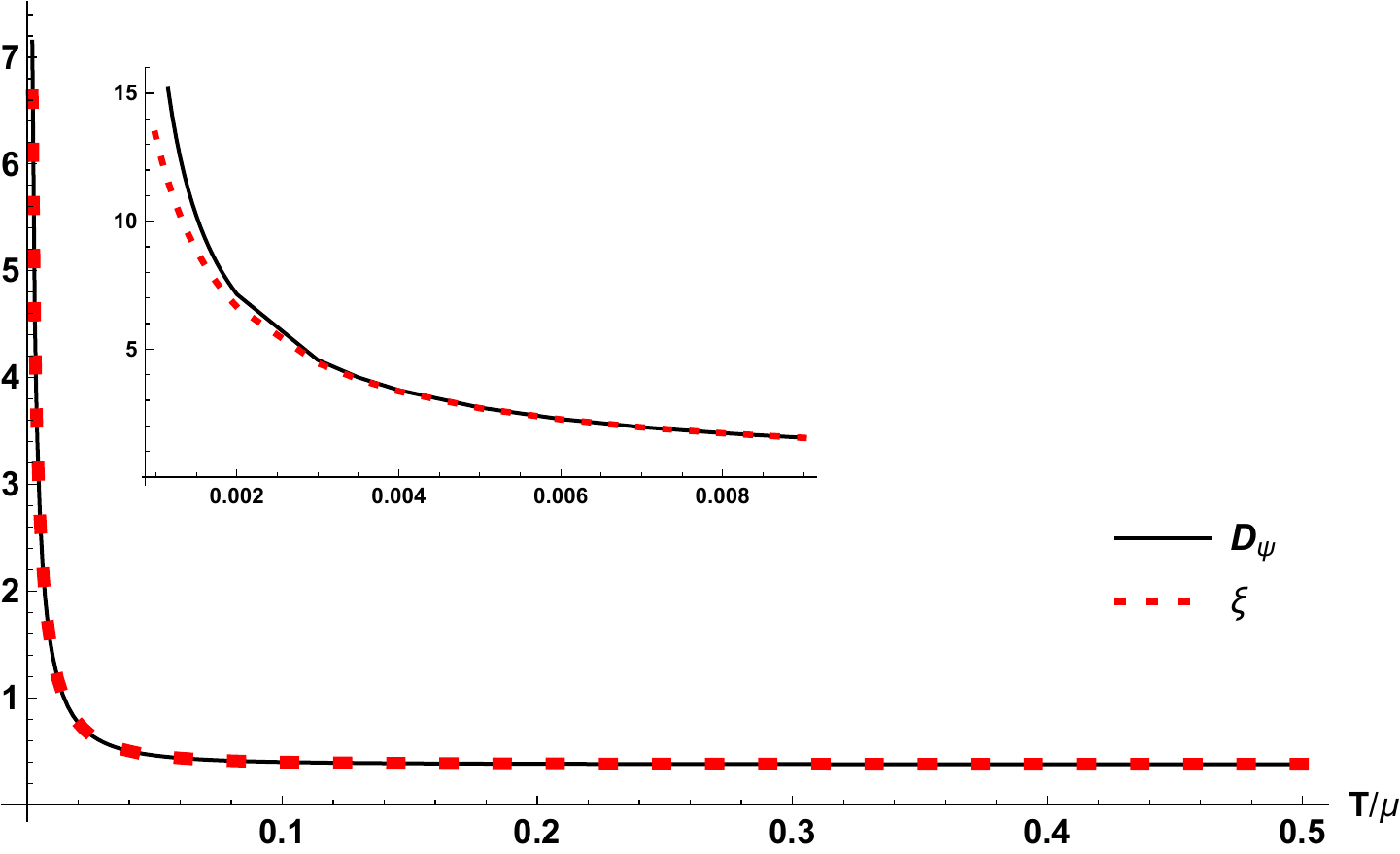}
\end{tabular}
\caption{Numerical check of equation \eqref{Dpsi=xi}.}
\label{fig:DpsiXivstemp}
\end{figure}

The presence of an overdamped mode when $\lambda\neq0$ is a consequence of the explicit breaking of the global shift symmetry $\psi\mapsto\psi+c$. It is easy to check that when $\lambda=0$, the gap vanishes and the mode \eqref{disprelpsik=0} becomes purely diffusive
\begin{equation}
\label{disprelpsik=0l=0}
\omega_{k=0,\lambda=0}=-i \xi q^2\,,\qquad \xi\equiv \frac{I_Y}{C_h Y_h}
\end{equation}
which coincides with the $k\to0$ limit of the CM diffusivity $G\Xi$ in equation \eqref{omegapm} of the main text. Turning $\lambda$ back on, we expect that 
\begin{equation}
\label{Dpsi=xi}
D_\psi=\xi+O(\lambda)=\frac{I_Y}{C_h Y_h}+O(\lambda)\,,
\end{equation}
which we verify numerically in figure \ref{fig:DpsiXivstemp}. Small deviations appear at very low temperature, upon approaching the pole collision. This is because the corrections of $O(\lambda)$ are no longer small compared to $T$.

\begin{figure}
\begin{tabular}{c}
\includegraphics[width=.75\textwidth]{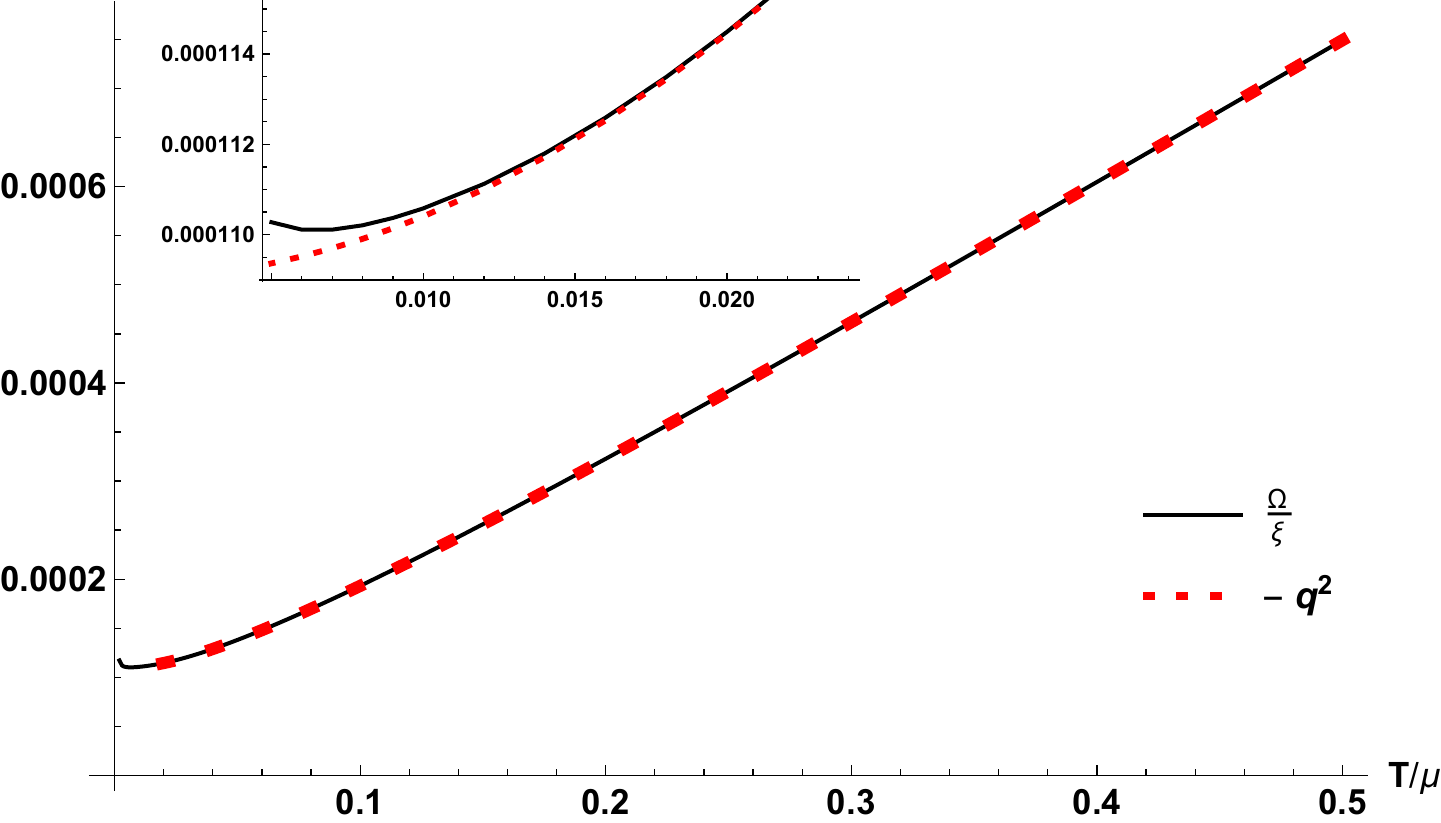}
\end{tabular}
\caption{Numerical check that there is a pole in \eqref{GRpsipsilambdanon0q} at $\omega=0$, $q^2=-m^2\simeq-\Omega/\xi$.}
\label{fig:OmegaOvXivstemp}
\end{figure}

Holographic and hydrodynamic correlators may differ by contact terms \cite{Kovtun:2012rj}, so we compare
\begin{equation}
\begin{split}
G^R_{\varphi_\perp\varphi_\perp}(\omega,q)-G^R_{\varphi_\perp\varphi_\perp}(\omega=0,q)=\frac{1}{G}\frac{i\omega}{(q^2+m^2)(i\omega-\Omega-q^2\xi)}+O(G^0,\Gamma^0\sim k^0)
\end{split}
\end{equation}
to
\begin{equation}
\begin{split}
G^R_{\delta\psi\delta\psi}(\omega,q)-G^R_{\delta\psi\delta\psi}(\omega=0,q)=
-\frac{i\omega \Omega\left(C_h Y_h\Omega+q^2(C_h Y_h D_\psi-I_Y)\right)}{\left(D_\psi q^2+\Omega\right)\left(i\omega-\Omega-D_\psi q^2\right)}\,.
\end{split}
\end{equation}
The two expressions can be seen to match since from \eqref{Dpsi=xi} $D_\psi\simeq\xi\simeq I_Y/(C_h Y_h)$ and $m^2\simeq\Omega/\xi$, as we verify in figure \ref{fig:OmegaOvXivstemp}, after identifying the boundary Goldstone mode and the vev of the bulk operator $\delta\psi$ as
\begin{equation}
\langle O_{\delta\psi}\rangle=\lambda^2\delta \psi^{(1)}=\frac{Gm^2}k\varphi\,.
\end{equation}

\section{$\lambda$ and $k$ dependence of the relaxation parameters\label{app:kandlscaling}}
\begin{figure}[!h]
\includegraphics[width=.47\textwidth]{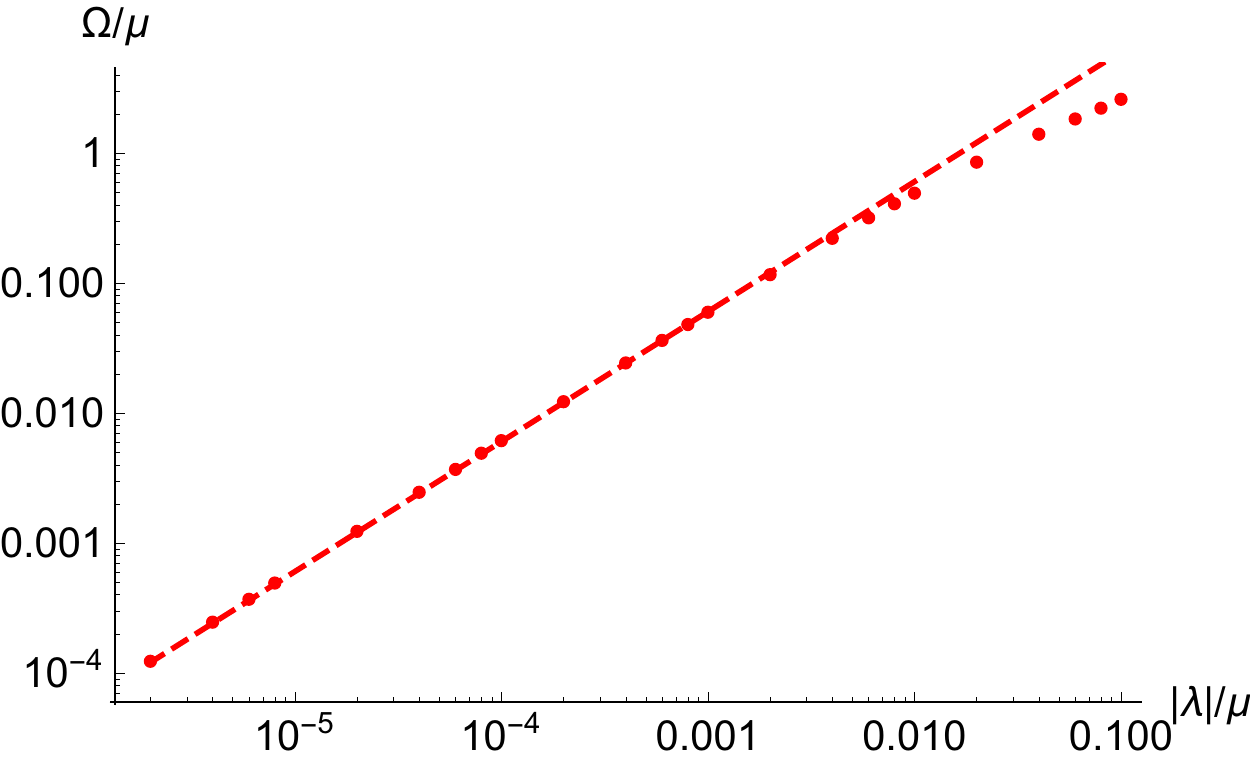}
\includegraphics[width=.5\textwidth]{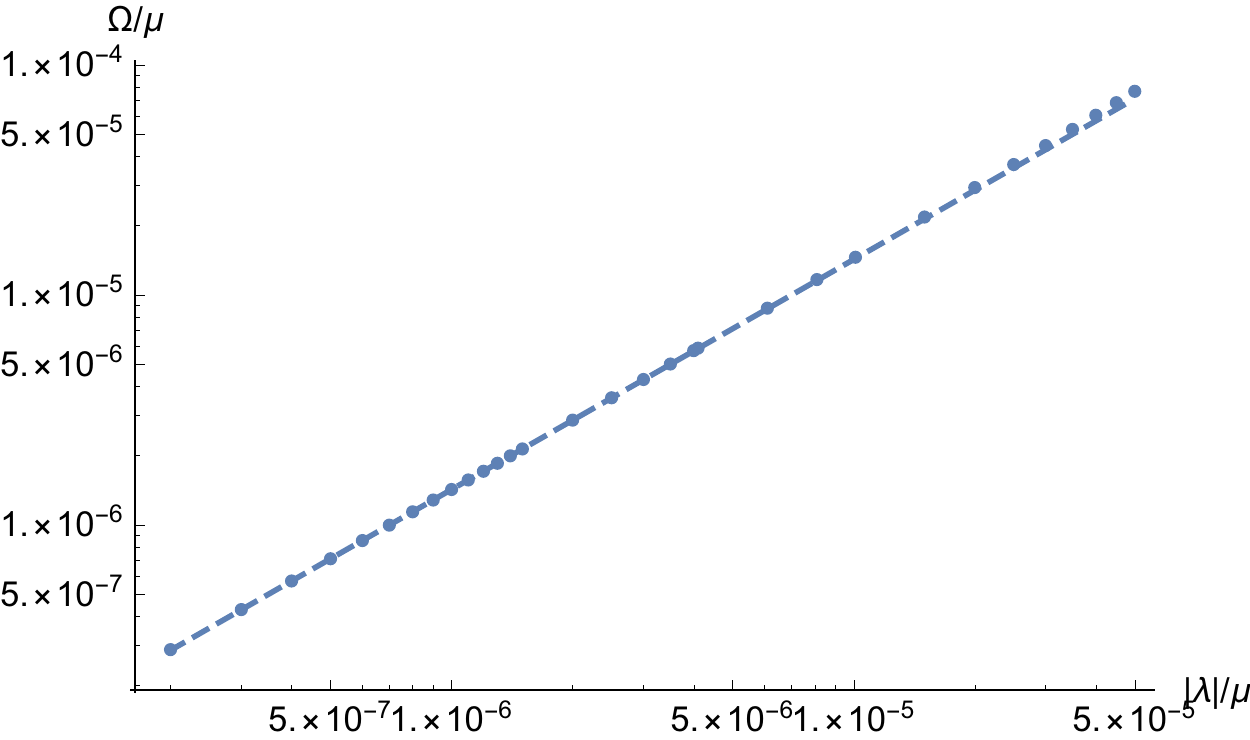}
\includegraphics[width=.47\textwidth]{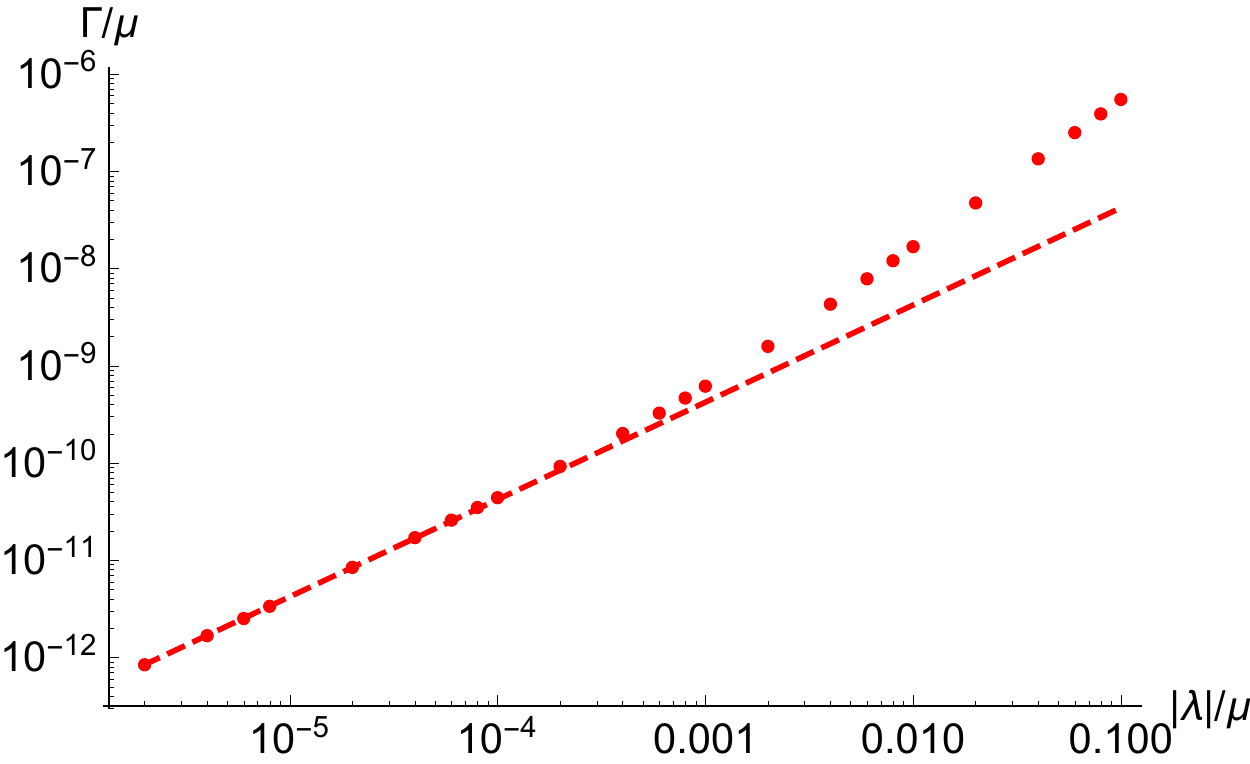}
\includegraphics[width=.5\textwidth]{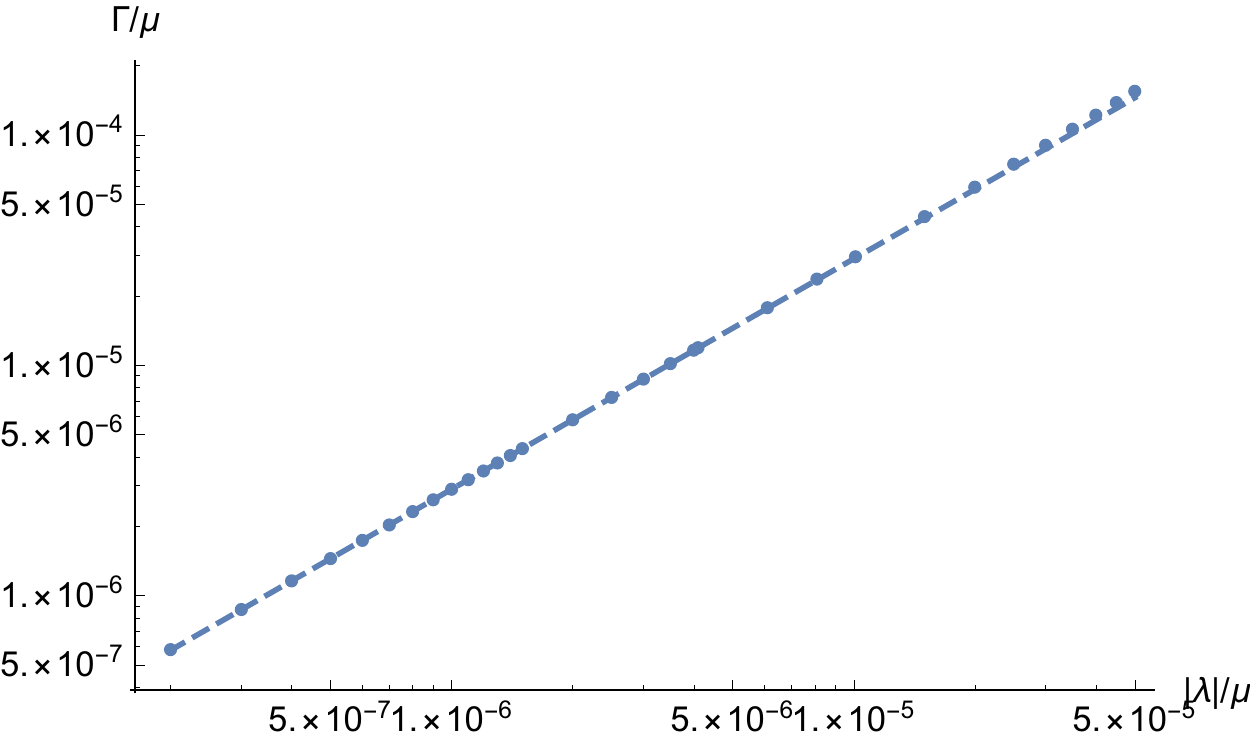}
\includegraphics[width=.47\textwidth]{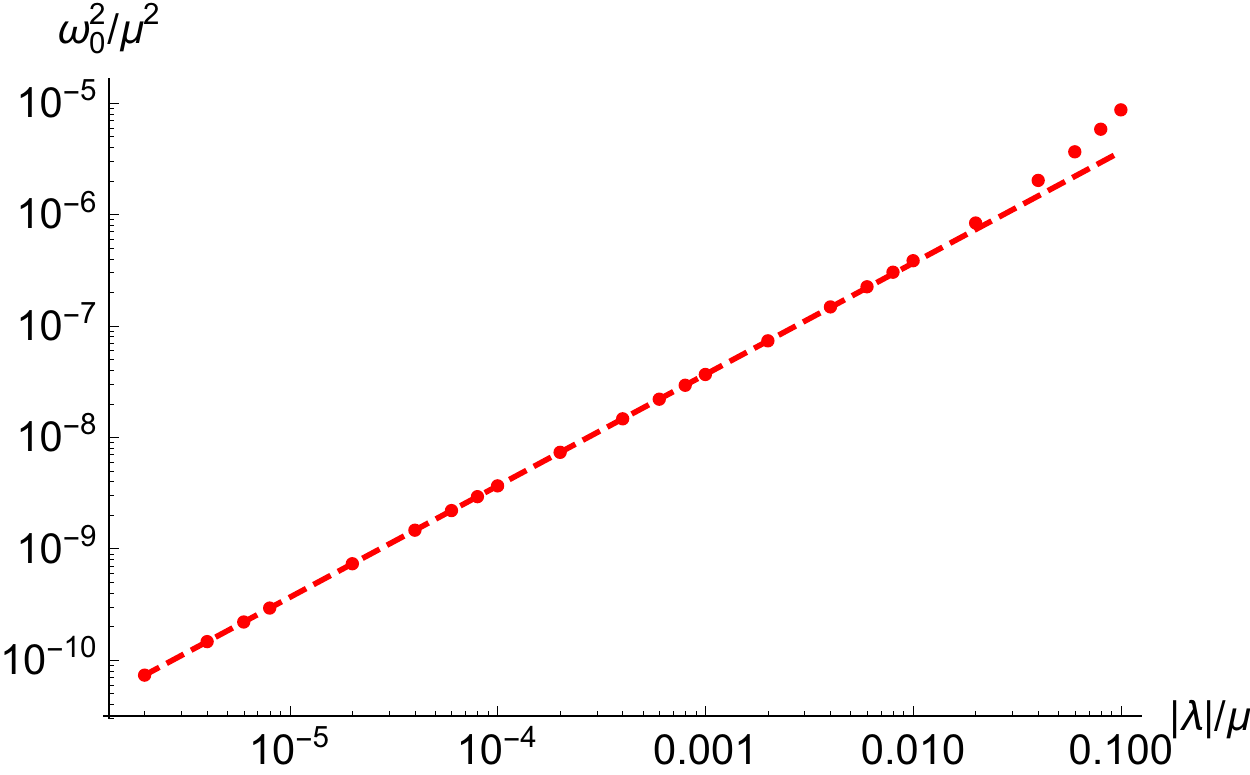}
\includegraphics[width=.5\textwidth]{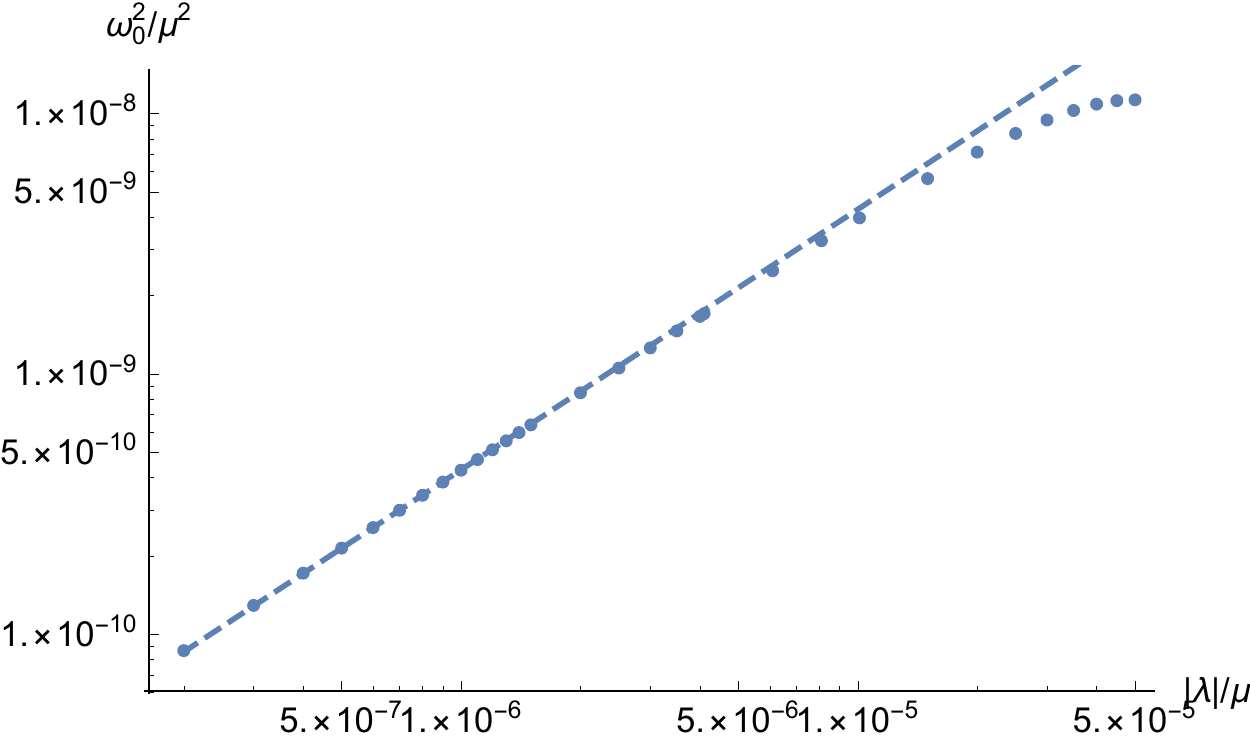}
\caption{Bi-log plots of the $\lambda$ dependence of $\Omega$, $\Gamma$ and $\omega_o$ and  their best fits (dashed lines) at $k/\mu=0.1$, revealing the scalings reported in equation \eqref{bestfitsrelaxparam}. The red dots are for $T/\mu=0.5$ while the blue dots are for $T/\mu=0.0035$.}
\label{fig:parvsl}
\end{figure}

\begin{figure}[!h]
\includegraphics[width=.47\textwidth]{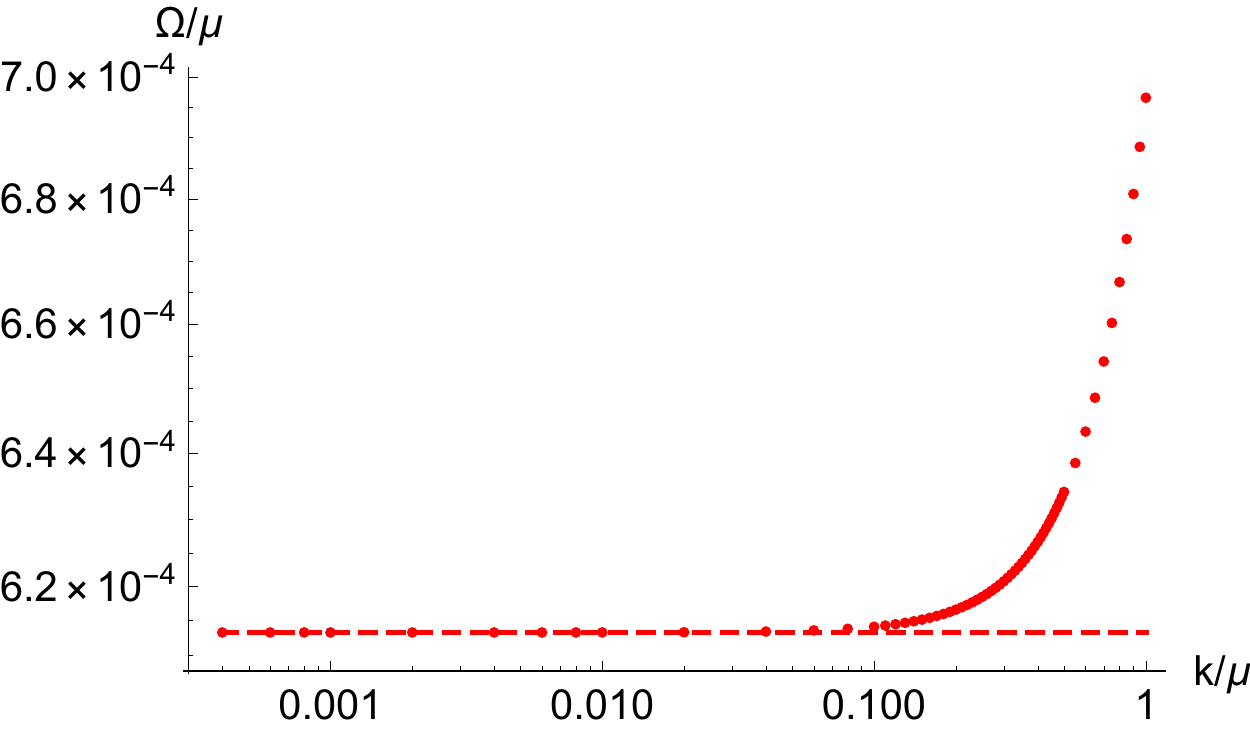}
\includegraphics[width=.5\textwidth]{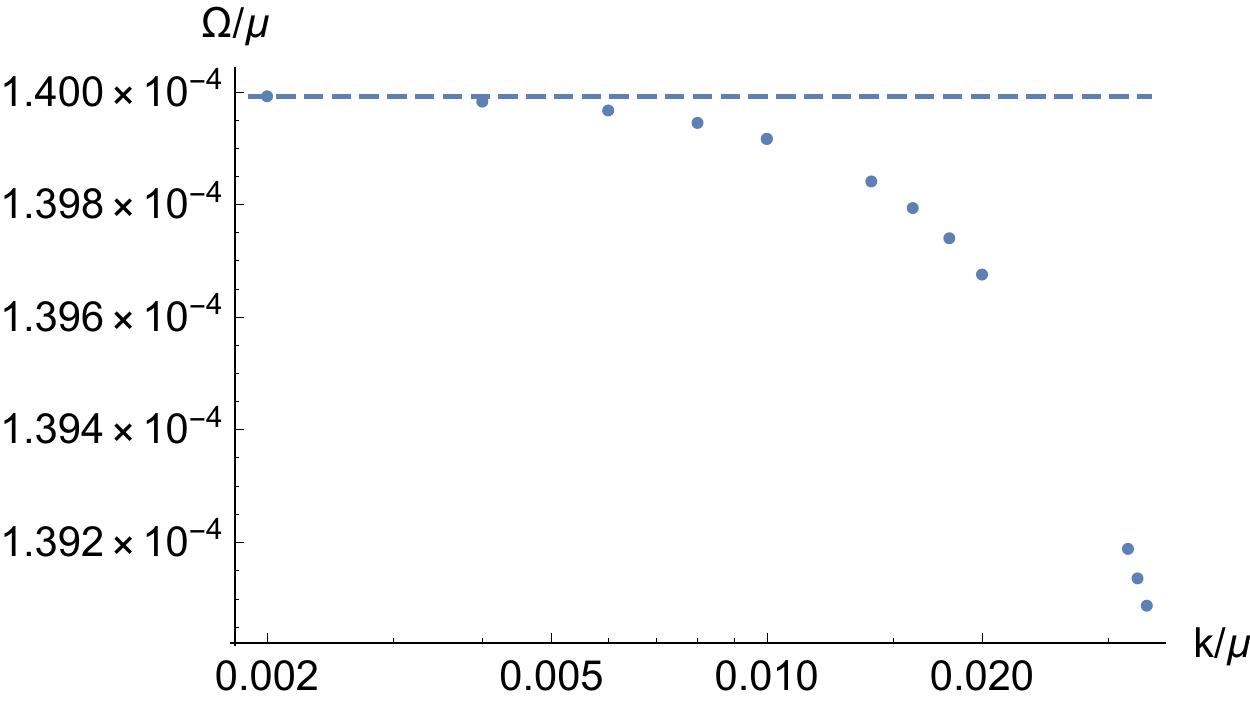}
\includegraphics[width=.47\textwidth]{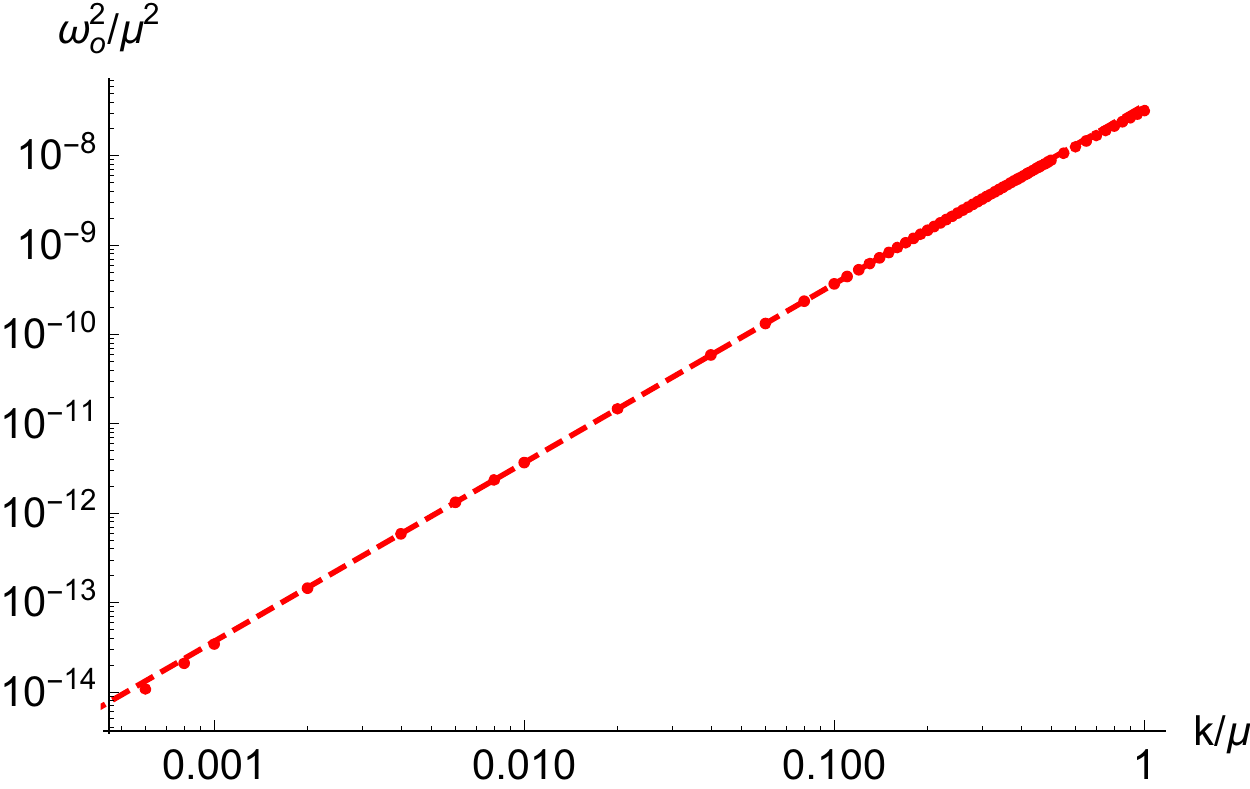}
\includegraphics[width=.5\textwidth]{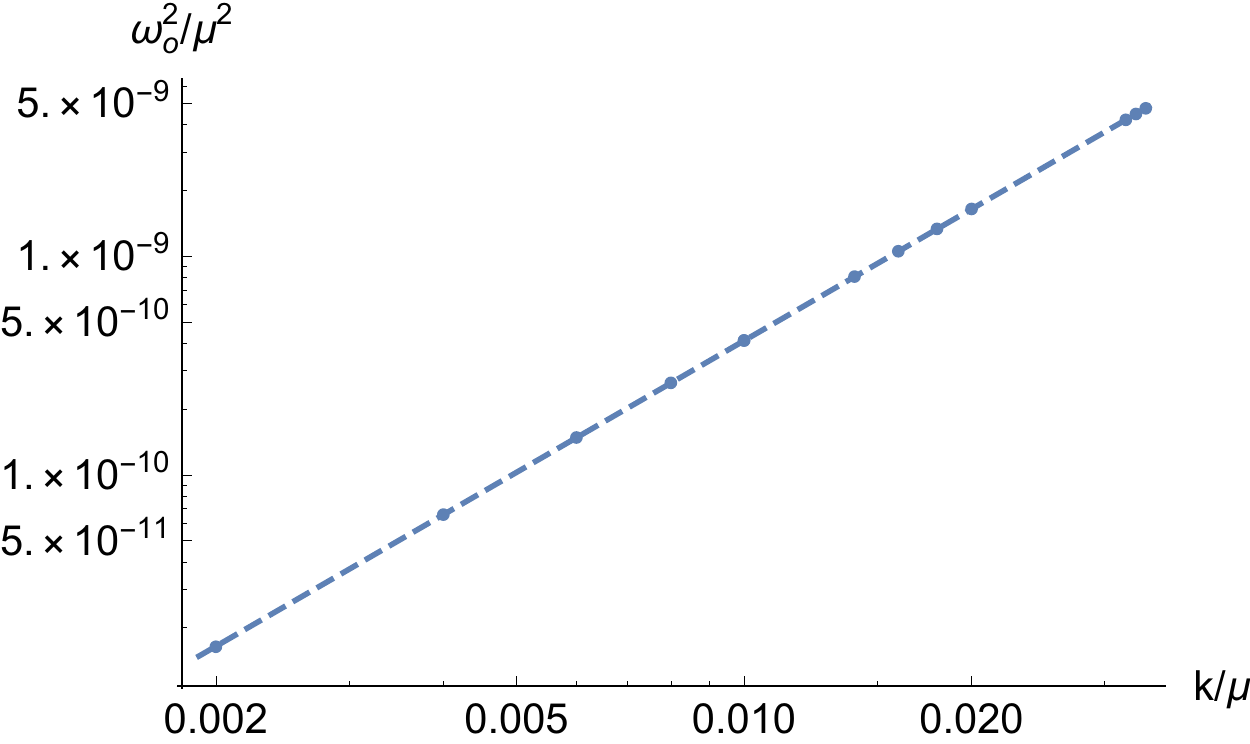}
\caption{Bi-log plots of the $k$ dependence of $\Omega$ and $\omega_o$ and their best fits (dashed lines) for $\lambda/\mu=-10^{-5}$, revealing the scalings reported in equation \eqref{bestfitsrelaxparam}. The red dots are for $T/\mu=0.5$ while the blue dots are for $T/\mu=0.0035$. As explained in the text, we have set $\Gamma=0$ for $T/\mu=0.0035$. For $\Omega$ (top row), the dashed line is a guide to the eye, from our lowest value of $k$ available, in very good agreement with the value of $\Omega$ at $k=0$ exactly.}
\label{fig:parvsk}
\end{figure}

\begin{figure}[!h]
\includegraphics[width=.47\textwidth]{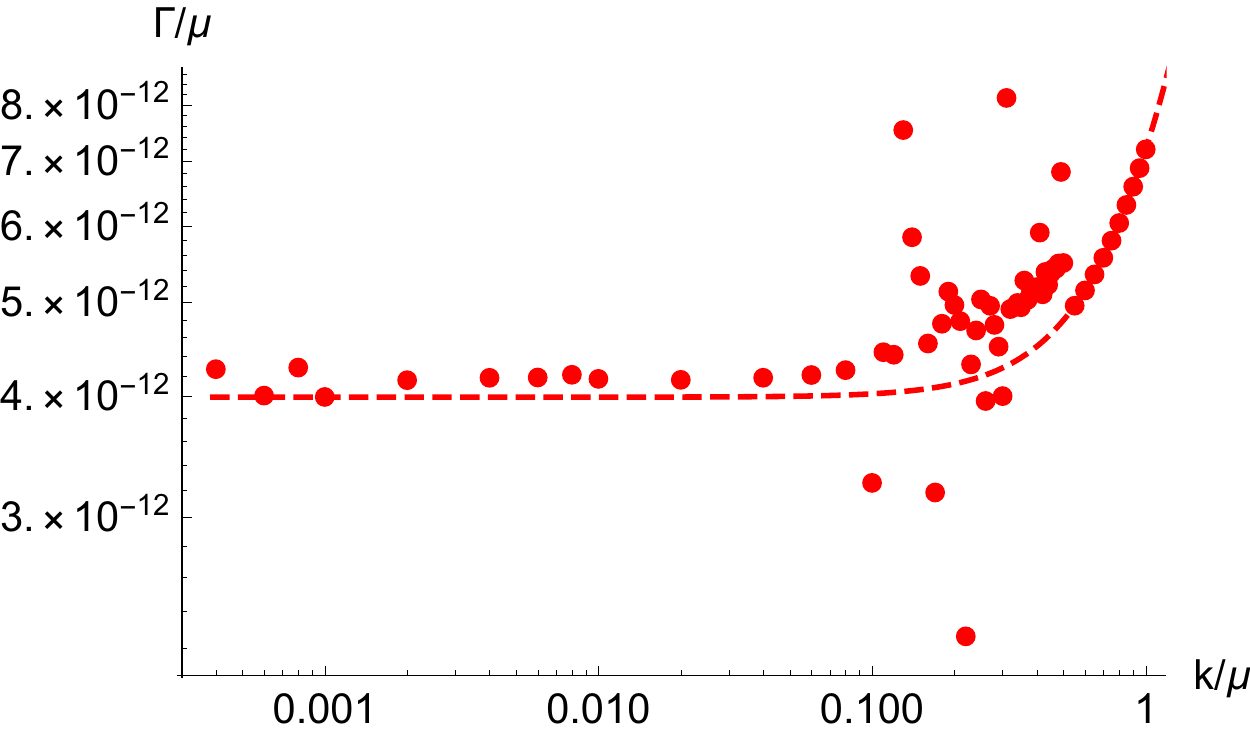}
\includegraphics[width=.47\textwidth]{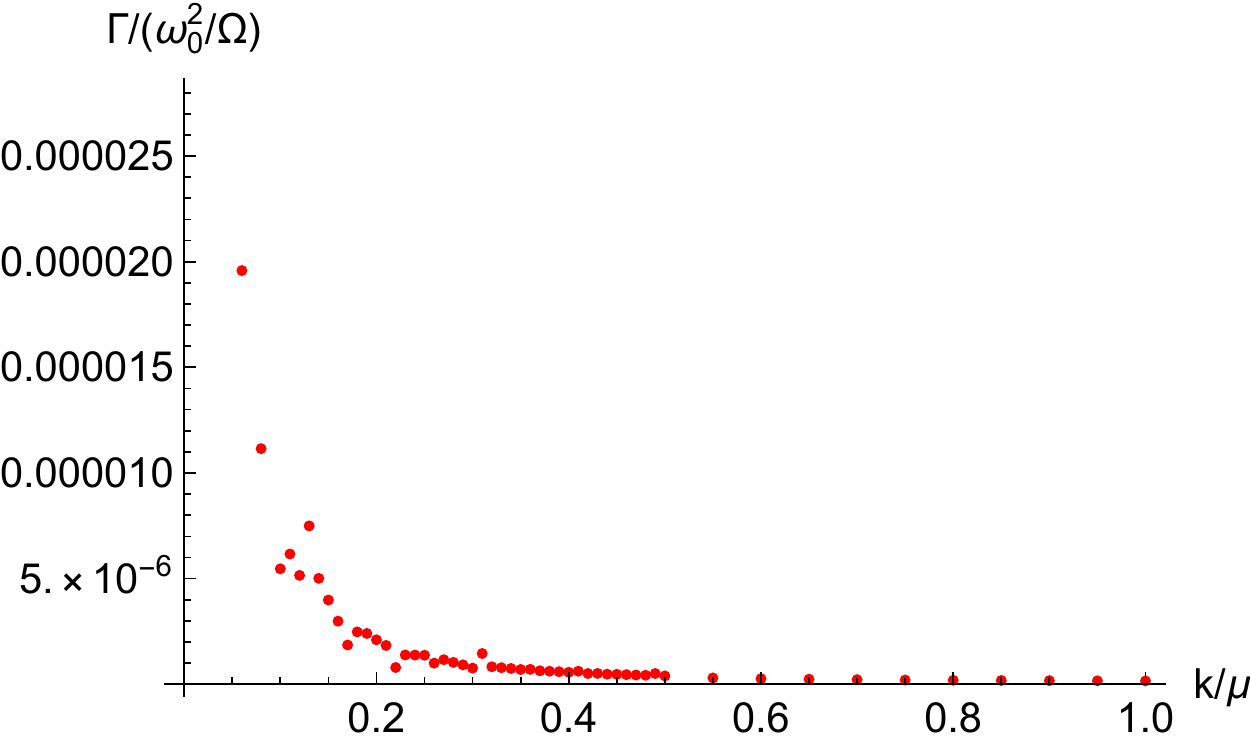}
\caption{ Left: $k$ dependence of $\Gamma$ for $T/\mu=0.5$ and $\lambda/\mu=-10^{-5}$, revealing the scaling reported in equation \eqref{bestfitsrelaxparam}.  Even though the data becomes noisy at low $k$, it is consistent with the best quadratic fit determined from the higher $k$ data. The tiny value for the intercept is close to the machine precision and we regard it as a numerical zero. 
Right: $k$ dependence of the  ratio $\Gamma/(\omega_o^2/\Omega)$ for $T/\mu=0.5$ and $\lambda/\mu=-10^{-5}$.}
\label{fig:Gammavskht}
\end{figure}

Using the quasi-normal modes computed numerically together with the hydrodynamic expression for the conductivity \eqref{hydrocondu} and the analytic expression for the diffusivity $\gamma_1$ (formula \eqref{gamma1xiUnrelaxed} of the main text), we have been able to compute the relaxation parameters $\Omega$, $\Gamma$ and $\omega_o$ and to fit their $k$ and $\lambda$ dependence. The results are shown in figures \ref{fig:parvsl}, \ref{fig:parvsk} and \ref{fig:Gammavskht} for two different values of the temperature: the red dots refer to $T/\mu=0.5$ and the blue dots to $T/\mu=0.0035$. 

The result of the fitting analysis is:
\begin{equation}
\label{bestfitsrelaxparam}
\Omega\sim |\lambda| k^0 \ , \qquad \Gamma \sim |\lambda| k^2 \ , \qquad \omega_o \sim |\lambda|^{1/2} k \ ,
\end{equation}
as reported in the main text. All our best fits include a small intercept, which is typically very close to our numerical zero and should be disregarded. 

At either very small $k$ or $\lambda$, the dc conductivity we use to extract the relaxation parameters becomes very large, and our numerical procedure loses accuracy. This is particularly apparent in the left panel of figure  \ref{fig:Gammavskht}, which shows the dependence of $\Gamma$ on $k$ at high $T$. We find values of $\Gamma/\mu$ of order $10^{-12}$, which we do not regard as very reliable. For all intents and purposes, $\Gamma$ should be set to zero, which we have done in the main text. This is further justified by the fact that in this range of $k$, $\Gamma\ll\omega_o^2/\Omega$, as shown in the right plot in fig \ref{fig:Gammavskht}. Moreover, we have verified that the values of $\Omega$ and $\omega_o$ obtained setting $\Gamma=0$ are quasi-identical to those obtained keeping $\Gamma\neq0$. At fixed $\lambda/\mu=-10^{-5}$ and $T/\mu=0.0035$, we obtain values for $\Gamma$ oscillating between tiny positive and negative values, once again with little effect on the values of $\Omega$ and $\omega_o$. For this reason, we have set $\Gamma=0$ for the results presented in the right column of figure \ref{fig:parvsk}. As a final check, we have verified that the $k\to0$ limits for the parameters $\Omega$ and $m$ agree with the values we find setting $k=0$ exactly, as is done in the previous section.

\begin{figure}
\includegraphics[width=.47\textwidth]{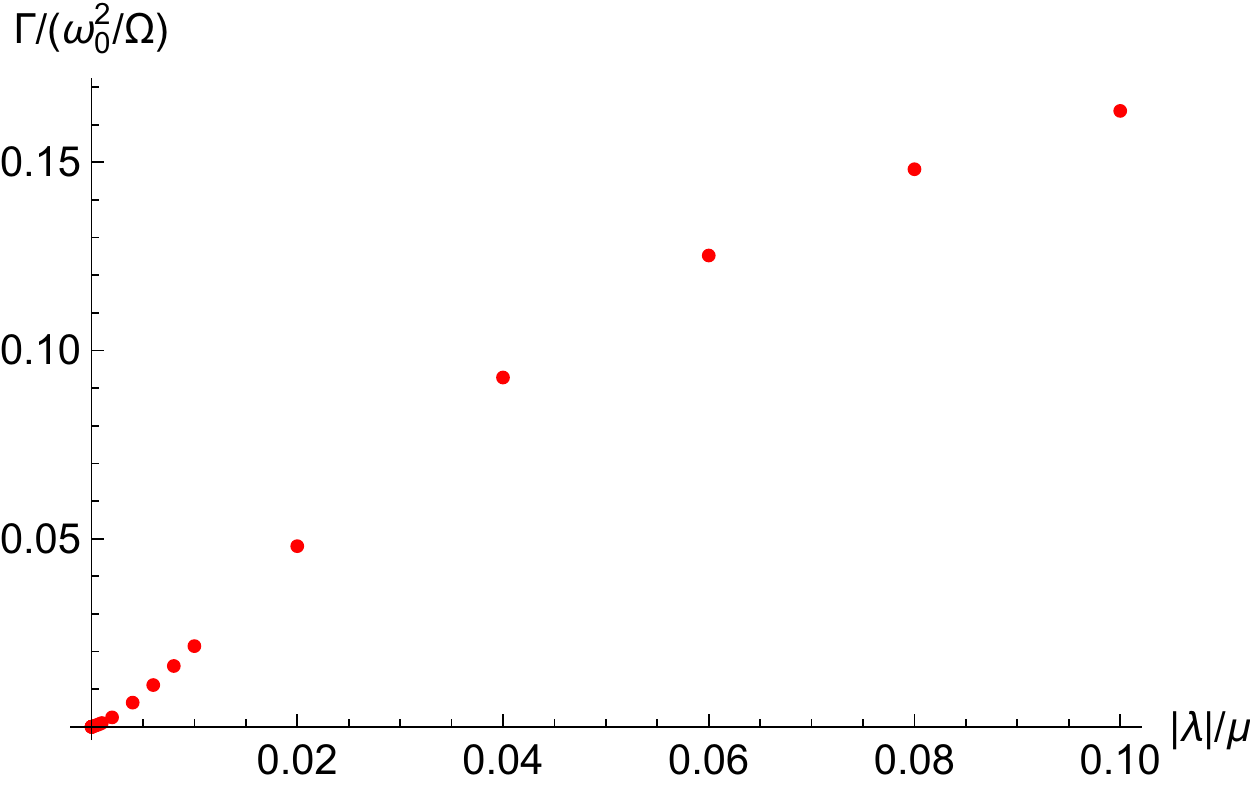}
\includegraphics[width=.5\textwidth]{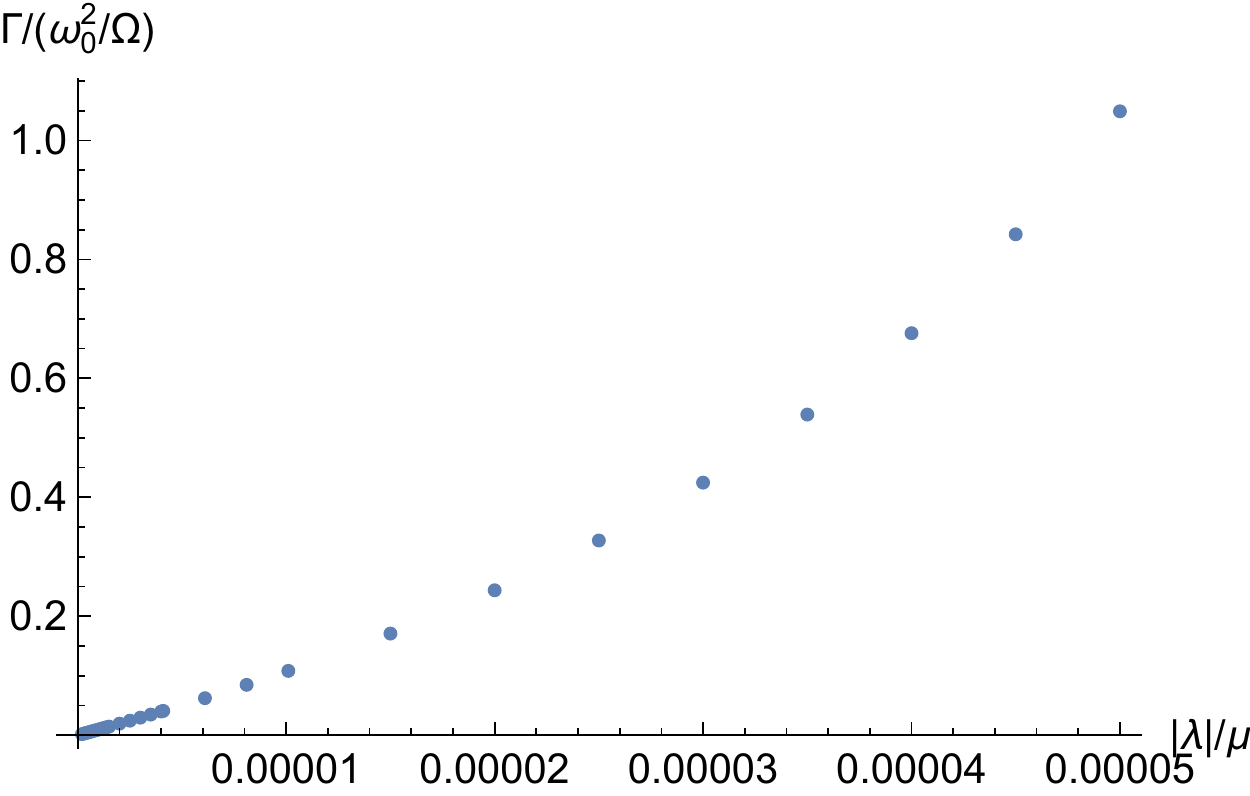}
\caption{$\lambda$ dependence of the quantity $\Gamma/(\omega_o^2/\Omega)$ for  $k/\mu=0.1$. Left:  $T/\mu=0.5$. 
	Right:  $T/\mu=0.0035$.
	For small enough $|\lambda|$, $\Gamma$ can be neglected compared to $\omega_o^2/\Omega$.}
\label{fig:ratiogammas}
\end{figure}

Figure \ref{fig:ratiogammas} shows the $\lambda$ dependence of the ratio $\Gamma/(\omega_o^2/\Omega)$ at high and low temperature. As one can see from the plots, for small $|\lambda|/\mu$ (eg $\lambda/\mu=-10^{-5}$ as it is considered in the main text) the ratio is very small with a very weak dependence on $|\lambda|/\mu$, implying that $\Gamma$ can be safely neglected as is done in presenting the results in the main text. The situation is different increasing $|\lambda|/\mu$, since $\Gamma$ becomes comparable to $\omega_o^2/\Omega$ for large enough $\lambda/\mu$. This is expected since by increasing $\lambda/\mu$ we are moving from the pseudo-spontaneous to the purely explicit regime, where any spontaneous component of the system is washed out and $\Gamma$ is the dominant relaxation scale.

\begin{figure}
\includegraphics[width=.47\textwidth]{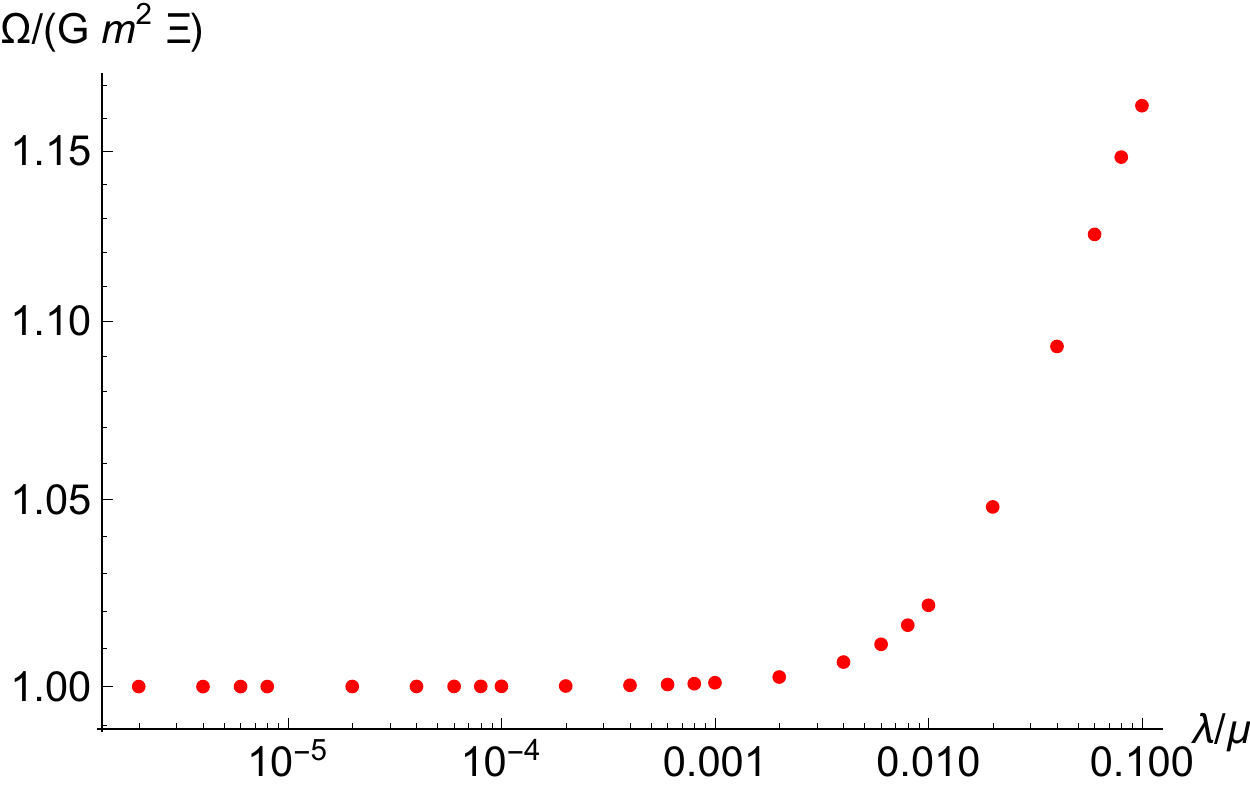}
\includegraphics[width=.5\textwidth]{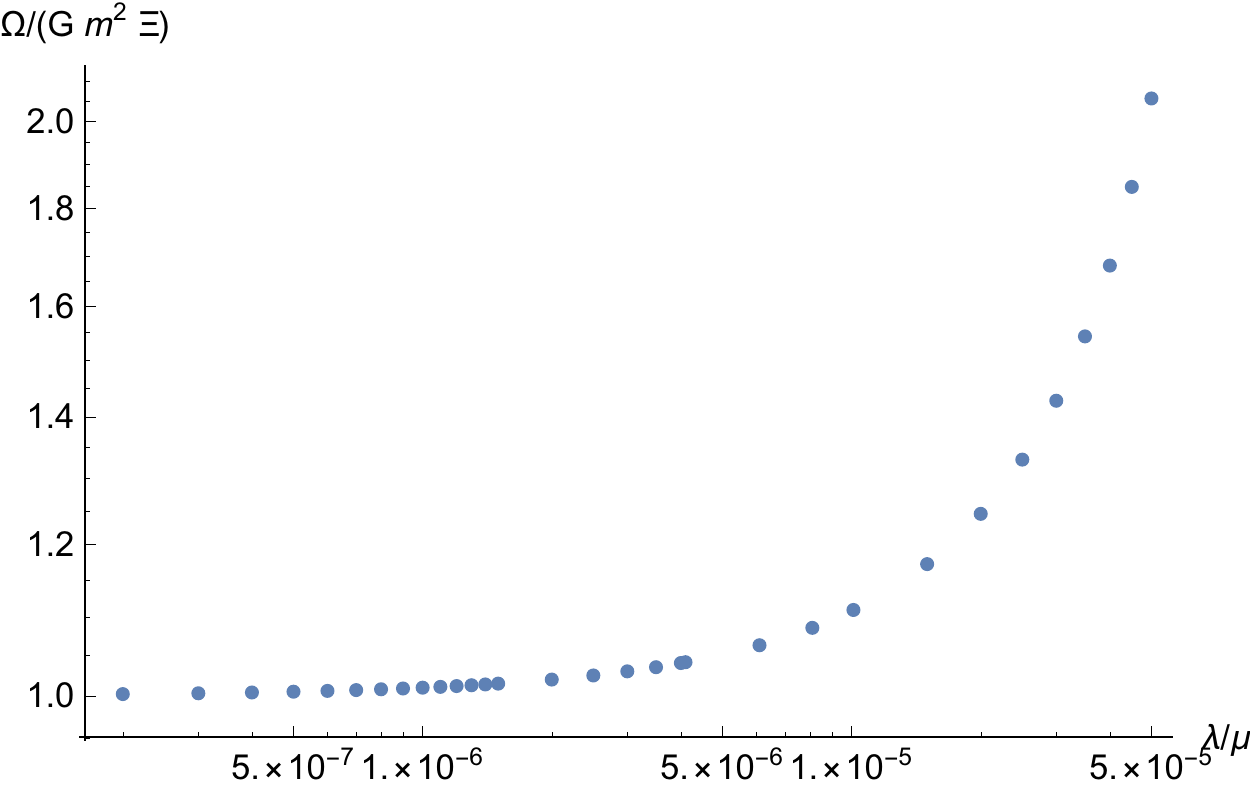}
\caption{$\lambda$ dependence of the ratio $\Omega/(G m^2 \Xi)$ for $k/\mu=0.1$. Left:  $T/\mu=0.5$. Right: $T/\mu=0.0035$.}
\label{fig:goldenratio}
\end{figure}

Figure \ref{fig:goldenratio} shows the $\lambda$ dependence of the ratio $\Omega/(G m^2\Xi)$ at high and low temperature. For small $|\lambda|$, the dependence on $|\lambda|$ is also very weak, before increasing more sharply as the explicit regime is approached.

Finally, we have checked that as $k$ is varied from very low values up to values $k/\mu=1$, the ratio $\Omega/(G m^2\Xi)$ shows no significant deviations from unity.

\begin{acknowledgments}
We are grateful to Hyun-Sik Jeong and Keun-Young Kim for pointing out some typos in a previous version of these appendices.
\end{acknowledgments}

\bibliography{PSTSB}

\end{document}